\documentclass[11pt]{article}
\pdfoutput=1
\usepackage{jcappub,natbib}
\input{colordvi.tex}

\def\ga{\mathrel{\raise.3ex\hbox{$>$\kern-.75em\lower1ex\hbox{$\sim$}}}}
\def\la{\mathrel{\raise.3ex\hbox{$<$\kern-.75em\lower1ex\hbox{$\sim$}}}}

\title{
Resonant particle production during inflation: a full analytical study
}

\author[a,b]{Lauren Pearce,}
\author[a,c]{Marco Peloso,} 
\author[d]{and Lorenzo Sorbo}

\affiliation[a]{School of Physics and Astronomy, University of Minnesota, Minneapolis, 55455 (USA)}
\affiliation[b]{Fine Theoretical Physics Institute, University of Minnesota, Minneapolis, 55455 (USA)}
\affiliation[c]{Minnesota Institute for Astrophysics, University of Minnesota, Minneapolis, 55455 (USA)}
\affiliation[d]{Amherst Center for Fundamental Interactions, Department of Physics, University of Massachusetts, Amherst, MA 01003 (USA)}

\abstract{We revisit the study of the phenomenology associated to a burst of particle production of a field whose mass is controlled by the inflaton field and vanishes at one given instance during inflation. This generates a bump in the correlators of the primordial scalar curvature. We provide a unified formalism to compute various effects that have been obtained in the literature and confirm that the dominant effects are due to the rescattering of the produced particles on the inflaton condensate. We improve over existing results (based on numerical fits) by providing exact analytic expressions for the shape and height of the bump, both in the power spectrum and the equilateral bispectrum. We then study the regime of validity of the perturbative computations of this signature. Finally, we extend these computations to the case of a burst of particle production in a sector coupled only gravitationally to the inflaton. 
}

\begin{document}

\begin{flushright} ACFI-T17-03, UMN--TH--3619/17, FTPI-MINN-17/03  \end{flushright}

\maketitle
\flushbottom

\section{Introduction} 
\label{sec:intro}

Inflation is typically described as a rather uneventful period in the history of the Universe -- an uneventfulness whose observable counterpart is the absence of features in the power spectrum of the primordial perturbations.  However, the time-dependence of the inflaton zero mode determines a ``clock'' which allows for phenomena that are localized in time. Such phenomena will generally induce features in the spectrum of metric perturbations, which might be observable in the Cosmic Microwave Background and in the Large Scale Structure data if they occurred $\sim 60$ to $\sim 50$ e-foldings before the end of inflation.

As is typical, features in spectra such as that of the CMB can be searched for more efficiently if templates describing them are available. It is thus important that the effects of motivated models that lead to such features receive a dedicated analysis, leading to (possibly analytical) expressions that make the dependence  on the parameters of the theory as transparent as possible.  Such features in the spectrum and bispectrum have been previously explored in the literature; see e.g.,~\cite{Adams:1997de,Lesgourgues:1999uc,Adams:2001vc,Kaloper:2003nv,Chen:2008wn,Chen:2009zp,Chen:2009we,Arroja:2011yu,Martin:2011sn,Adshead:2011jq}  for a (incomplete) list of references that studied the effects of phase transitions during inflation or features in the inflaton potential.  In this paper we analyze features in the spectrum and bispectrum in models  where a sudden event of particle production occurs during inflation.

In this scenario, initially proposed in~\cite{Chung:1999ve}, the rolling zero mode of the inflaton $\phi$ controls the mass of an auxiliary field $\chi$.  We will parametrize, as is usual, the $\phi$-dependent mass of $\chi$ as $m_\chi=g\,\left(\phi-\phi_*\right)$, with $g$ and $\phi_*$ constants.  If, during the rolling of $\phi$, $m_\chi$ crosses zero quickly enough, quanta of $\chi$ are explosively created.  The presence of those quanta can lead to observable signatures  {\em (i)}  by  backreacting on the evolution of the zero mode of the inflaton, $\varphi \left( t \right)$, which in its turn affects the way vacuum fluctuations of $\phi$ are amplified by the time-dependent background, {\em (ii)} by rescattering  off the zero mode of the inflaton, producing inflaton quanta, through a $\varphi \left( t \right) \delta \phi \, \chi^2$ coupling that arises from the above mass term, and {\em (iii)} by four point interactions mediated by the vertex $\delta \phi^2 \, \chi^2$ which also arises from the mass term. 

Several consequences of this coupling were explored in~\cite{applications}. In particular, we note that a sequence of $g^2 \left( \phi - \phi_{*,i} \right)^2 \chi_i^2/2$ couplings is used in trapped inflation~\cite{Green:2009ds}. Trapped inflation is a specific realization of the idea that the inflaton kinetic energy can be dissipated through coupling the inflaton to auxiliary fields, which was first proposed in the context of warm inflation~\cite{Berera:1995ie,Berera:1998px,Graham:2008vu,BasteroGil:2009ec,Bartrum:2013fia,Bastero-Gil:2014raa}.  The spectrum and bispectrum in trapped inflation were originally analyzed in~\cite{Green:2009ds}, with the bispectrum further studied in~\cite{LopezNacir:2011kk} (see also \cite{other-trapped}). We have recently recomputed the primordial spectrum and bispectrum in trapped inflation, improving over some crucial approximations made in previous studies of that model regarding the correlators of the fields sourcing the inflaton perturbations~\cite{Pearce:2016qtn}.   We note that while the operators of trapped inflation have the same form as the coupling considered here, the phenomenology of trapped inflation is more complicated as it is computed in the regime where there is a strong backreaction from the produced particles on the background field.  As an example, the sourced contribution to the inflaton perturbation dominates over the vacuum contribution, which is contrary to the limit considered in this work, due to phenomenological constraints. 

In this work we also consider, for the first time, a variant of the scenario described above, in which particle production occurs in a hidden sector, similar to what was proposed in~\cite{Barnaby:2012xt}.  This is a three field model, in which in addition to the inflaton field $\varphi$ the scalar fields $\psi$ and $\chi$ are present with the coupling $g^2(\psi - \psi_*)^2\,\chi^2 \slash 2$.  Although not the inflaton, the $\psi$ field will have an evolving expectation value for some period during inflation, which we assume passes through $\psi_*$, generating explosive production of $\chi$ quanta.  The hidden $\left\{ \psi ,\, \chi \right\}$ sector is coupled to the visible sector (including the inflaton) gravitationally.  Consequently, the same three processes (\textit{i})-(\textit{iii}) mentioned above are also present in this construction, albeit with slightly different vertices.

In this work we study the processes (\textit{i})-(\textit{iii}) in a unified manner by making use of the in-in formalism. This allows us to present analytical formulae for the corrections they induce to the two- and three-point function of the primordial scalar perturbations.  These correlators turn out to depend on two parameters. The first and obvious one is the coupling $g$, which is limited by perturbativity. The second is the combination $g\,\dot\varphi_*/H^2$ where $H$ is the Hubble rate during inflation and $\dot\varphi_*$ is the velocity of the inflaton at the time of particle production. Nonadiabaticity (which is necessary for the production of $\chi$ quanta to be efficient enough) requires this second parameter to be much larger than unity. The presence of this large parameter allows us to determine that the rescattering process {\em (ii)} gives the dominant contribution to the feature in the metric perturbation spectrum, which we denote as $\delta_{\zeta,1}(k)$ and provide in analytical form in eqs.~(\ref{P-int-res}) and~(\ref{shapes}). This same process gives also the dominant contribution to the feature in the bispectrum, which we discuss in Section~\ref{subsec:BS}. To our knowledge, the precise analytic form of these features was not given in the previous literature.

Since the theory contains a small expansion parameter $g$ along with a large one, $g\,\dot\phi_*/H^2$, it is not trivial to determine the region of validity of the perturbative regime. As we discuss in Appendix~\ref{app:Feynman}, the use of the in-in formalism allows us to determine a set of rules for the scaling with $g$ and $g\,\dot\phi_*/H^2$ of the various diagrams. We show, for the first time, that perturbativity requires $g^2\lesssim 3$ as the result of a nontrivial competition between powers of $2 \pi$ and the quantity $\sqrt{\vert \dot{\varphi}\vert}/H$, which is determined by the normalization of the unperturbed scalar power spectrum.

Various aspects of the phenomenology of this model have been considered in the past. The works~\cite{Chung:1999ve,Elgaroy:2003hp,Romano:2008rr} have discussed the effects of the process {\em (i)} (the correction to the evolution of the zero mode of the inflaton) on the spectrum of perturbations. As we show in Appendix~\ref{app:dzeta}, this effect is subdominant with respect to the effect {\em (ii)} of rescattering of quanta of $\chi$ on the zero mode of the inflaton, which has been studied  in a series of papers~\cite{Barnaby:2009mc,Barnaby:2009dd,Barnaby:2010ke} where  numerical results and fits were provided. Our analytical results are close to, but do not always agree with,  the results of these works. For instance, reference~\cite{Barnaby:2009dd} finds that the amplitude of the feature in the power spectrum goes as $g^{15/4}$ while our analytical result gives an amplitude proportional to $g^{7/2}$. Moreover, our analytic results show that, at momenta greater than those of  the peak, the correction to the power spectrum is characterized by oscillations, modulated by an amplitude that decreases as $k^{-3}$, while those previous works described it with a monotonic and  exponentially decreasing function. We attribute these differences to the use of numerical fits used in these papers.  Let us finally note that the effect {\em (iii)}, which we find to give a contribution that is subdominant with respect to {\em (ii)}, has never been studied in the literature. ~\footnote{An analogous  $4-$point  $\delta g^2 \, \chi^2$ interaction between quanta of $\chi$ and tensor metric perturbations $\delta g$  was  discussed by~\cite{Carney:2012pk}, which studied the corrections to the spectrum of tensors in this model also using the in-in formalism.  The tensor mode produced in this model was also studied in ~\cite{Cook:2011hg,Senatore:2011sp,Carney:2012pk,Barnaby:2012xt} and it is typically small due to the non-relativistic nature of the sourcing $\chi-$quanta; for this reason, we do not discuss the tensor production in this work.} 

The plan of this paper is the following. In Section \ref{sec:main} we compute the modifications of the inflation power spectrum and bispectrum in the case in which the field $\chi$ is directly coupled to the inflaton field and its nonperturbative production is due to the inflaton motion. Several details are worked out in a number of appendices. While in Appendix \ref{app:PS} we perform the exact computations of the dominant diagrams that lead to these modifications,  in Appendix \ref{app:Feynman} we provide Feynman rules that allow one to estimate the contribution of  all diagrams and to study the limit mentioned above on the validity of perturbation theory. Another important appendix is  Appendix~\ref{app:dg}, where we show that gravitationally mediated interactions are subdominant in this model and that metric perturbations can be ignored (in a specific gauge; see below for technicalities). In Section \ref{sec:hidden} we then study the modifications of the inflaton power spectrum in the case in which the particle production occurs in a sector that is only gravitationally coupled to the inflaton. In this case, including metric perturbations is key to obtaining the signature. In Section \ref{sec:discussion} we summarize our findings and conclude.

\section{Isolated sudden particle production during inflation  }
\label{sec:main}

We consider the potential 
\begin{equation}
V \left( \phi \right) + \frac{g^2}{2} \left( \phi - \phi_* \right)^2\,\chi^2 \;, 
\end{equation}
where $\phi$ is the inflaton and $\chi$ is a field coupled to it. The quantity $\phi_*$ is a given value that is crossed by the inflaton during inflation. We denote by $\tau_*$ the conformal time at which this happens.  At this time the mass of $\chi$ varies non-adiabatically, resulting in copious production of quanta of $\chi$. 

To perform computations, we separate the lagrangian of the model into unperturbed plus perturbed terms as follows,
\begin{eqnarray} 
S &=& \int d^4 x \, a^4 \, \left[ {\cal L}_{0} +  {\cal L}_{\rm int} \right] \;\;, \nonumber\\ 
{\cal L}_0 &\equiv& \frac{1}{2 \, a^2}  \phi'{}^{2} - \frac{1}{2 \, a^2}  \partial_i \phi^2   +  \frac{1}{2 \, a^2}  \chi'{}^{2} -  \frac{1}{2 \, a^2}  \partial_i \chi^2 
- V \left( \phi \right) - \frac{g^2}{2} \left( \varphi - \phi_* \right)^2 \chi^2 \;\;, \nonumber\\ 
{\cal L}_{\rm int} &\equiv&  - g^2 \left( \varphi - \phi_* \right) \delta \phi \, \chi^2 - \frac{g^2}{2} \delta \phi^2 \, \chi^2  \;\;, 
\label{action}
\end{eqnarray} 
where a prime denotes a derivative with respect to conformal time $\tau$, and the index $i=1,2,3$ runs over the spatial coordinates.  In these equations we have chosen the line element $d s^2 = a^2 \left( \tau  \right) \left[ - d \tau^2 + d x^i \, d x^i \right]$, disregarding  metric perturbations. We show in Appendix \ref{app:dg} that this is a good approximation for the purpose of computing the corrections to the scalar curvature power spectrum due to the $\chi$ production. We have then split the inflaton field as 
\begin{equation}
\phi \left( t ,\, \vec{x} \right) = \varphi \left( t \right) + \delta \phi \left( t ,\, \vec{x} \right) \;\;,\;\; \varphi \equiv \left\langle \phi \right\rangle \Big\vert_{g=0} \;, 
\label{varphi}
\end{equation} 
namely, $\varphi$ is the homogeneous inflaton field in absence of its interactions with the field $\chi$. We note that $\delta \phi$ contains both a homogeneous and non-homogeneous component, where the former encodes the effect of the back-reaction of the quanta $\chi$ on the homogeneous inflaton.  

Our goal is to compute the power spectrum of  the gauge invariant curvature perturbation $\zeta$, defined through 
\begin{equation}
\left\langle \zeta \left( \tau ,\, \vec{k} \right) \,  \zeta \left( \tau ,\, \vec{k}' \right) \right\rangle \equiv \frac{2 \pi^2}{k^3} \, P_\zeta \left( k \right) \, \delta^{(3)} \left( \vec{k} + \vec{k}' \right) \;,
\end{equation} 
and the analogously-defined bispectrum.  The quantity $\zeta$ is defined in Appendix \ref{app:zeta}, where we show that in the super-horizon regime we can accurately approximate it as 
\begin{equation}
\zeta \left( \tau ,\, \vec{k} \right) \simeq - \frac{\phi' \left( \tau \right)}{{\cal H} \left( \tau \right)} \; \delta \phi \left( \tau ,\, \vec{k} \right) \;\;\;,\;\;\; k \ll {\cal H} \;, 
\label{zeta-main} 
\end{equation}
where ${\cal H} = a \,H = \frac{a'}{a}$ is the comoving Hubble rate. Therefore in the main text we compute the two-point and three-point expectation values of the inflaton perturbations, treating the expression (\ref{zeta-main}) as an equality, with the understanding that it is evaluated in the super-horizon regime. 

We denote by $\phi^{(0)} \left( t ,\, \vec{x} \right) \equiv \varphi \left( t \right) + \delta \phi^{(0)} \left( t ,\, \vec{x} \right)$ and $\chi^{(0)} \left( t ,\, \vec{x} \right)$ the inflaton and the $\chi-$field at zeroth order in the interaction lagrangian ${\cal L}_{\rm int}$.  From these fields we can compute the zeroth-order power spectrum 
\begin{equation}
P_\zeta^{(0)} \left( \tau,\,  k \right)  \Big\vert_{\tau \gg \tau_* ,\, k \ll {\cal H}}  = \frac{k^3}{2 \pi^2} \, \frac{\varphi'{}^2 \left( \tau \right)}{{\cal H}^2 \left( \tau \right)} \, \left\langle \delta \phi^{(0)} \left( \tau ,\, \vec{k} \right)  \delta \phi^{(0)} \left( \tau ,\, \vec{k}' \right) \right\rangle'  \;, 
\label{P-0} 
\end{equation} 
(where prime on an expectation value denotes the expectation value without the corresponding Dirac $\delta-$function), as well as the perturbative corrections to it, using the in-in formalism 
\begin{eqnarray}
\delta P_\zeta \left( \tau ,\, k \right) \Big\vert_{\tau \gg \tau_* ,\, k \ll {\cal H}} &= & \frac{k^3}{2 \pi^2} \, \frac{\varphi'{}^2  \left( \tau \right)}{{\cal H}^2 \left( \tau \right)} \,  \sum_{N=1}^{\infty} \left( - i \right)^N \, \int^\tau d \tau_1 \dots \, \int^{\tau_{N-1}} d \tau_N \nonumber\\ 
& & \quad\quad \times \left\langle 
\left[ 
\left[ 
\cdots  
\left[ 
\delta \phi^{(0)} \left( \tau ,\, \vec{k} \right)  \delta \phi^{(0)} \left( \tau ,\, \vec{k}' \right)  ,\, H_{\rm int}^{(0)} \left( \tau_1 \right) 
\right] , \cdots 
\right] ,\,  H_{\rm int}^{(0)} \left( \tau_N \right) 
\right] 
\right\rangle' \;, \nonumber\\ 
\label{P-int} 
\end{eqnarray} 
where the (unperturbed) interaction hamiltonian is given by 
\begin{equation}
H_{\rm int}^{(0)} \left( \tau_i \right) \equiv - a^4 \left( \tau_i \right) \, \int d^3 x \; {\cal L}_{\rm int} \left( \phi^{(0)} \left( \tau_i ,\, \vec{x} \right) ,\,  \chi^{(0)} \left( \tau_i ,\, \vec{x} \right) \right) \;. 
\label{Hint}
\end{equation}  
A similar expansion in the in-in formalism applies to the bispectrum, involving three $\delta \phi^{(0)}$ operators.

We note that the prefactor $ \frac{\varphi'{}^2  \left( \tau \right)}{{\cal H}^2 \left( \tau \right)} $ in the expressions (\ref{P-0}) and (\ref{P-int}) is understood as the ratio between unperturbed background quantities, namely disregarding the backreaction of the produced quanta of $\chi$ on the evolution of the inflaton and of the scale factor. In principle, one should also account for the perturbative corrections to this prefactor. However, this can be  disregarded, for the following reason: in the in-in formalism computation we evaluate the expressions  (\ref{P-0}) and (\ref{P-int}) in the super-horizon regime, well after the production of $\chi$ has taken place. As we show below, after production the energy density of the quanta of $\chi$ redshifts essentially as that of matter, so it soon becomes negligible with respect to that of the inflaton. Therefore, at $\tau \gg \tau_*$ the field $\chi$ has redshifted away, and the inflaton and the scale factor evolve according to the standard attractor solution, dictated simply by the potential $V \left( \phi \right)$. The departure from this attractor solution in a neighborhood of $\tau_*$ only results in a overall time shift of the background solution.  Namely at $\tau \gg \tau_*$ we simply have $\langle \phi \left( \tau \right) \rangle = \langle \varphi \left( \tau + \Delta \tau \right) \rangle $ (and analogously for the scale factor), where  $\Delta \tau $ is a fixed (namely, $k-$independent) quantity that does not affect the scale dependence of the power spectrum. It only changes the relation between the scale $k$ and the number of e-folds before the end of inflation when this scale left the horizon.   Consequently, the prefactor takes the form given above,  in which only unperturbed quantities appear. 

We present the computation of the power spectrum and bispectrum in three subsections. In  Subsection \ref{subsec:zero} we first calculate the unperturbed fields $\phi^{(0)}$ and $\chi^{(0)}$, and additionally we evaluate the unperturbed expression (\ref{P-0}). In  Subsection \ref{subsec:int} we compute the dominant  correction to the power spectrum contained in (\ref{P-int}). This provides the dominant phenomenological signature of the particle production.  Specifically, we compute the terms that can be diagrammatically understood as the one-loop processes in Figure \ref {fig:diagrams}. In Appendix \ref{app:Feynman} we study for what range of the coupling $g$ our result is under perturbative control.  Finally, we address the bispectrum in Subsection \ref{subsec:BS}.

\subsection{Power spectrum to zeroth order in ${\cal L}_{\rm int}$} 
\label{subsec:zero}

In this subsection we study the zeroth order fields described by the unperturbed lagrangian ${\cal L}_0$ in eq. (\ref{action}). This lagrangian essentially controls three quantities: 

(i) The unperturbed background inflaton, which evolves according to 
\begin{equation} 
 \varphi''   + 2 \frac{a'}{a} \, \varphi'  + \frac{d V \left( \phi \right)}{d \phi} \Bigg\vert_{\phi =  \varphi   } = 0 \;,  
\end{equation}
in absence of any backreaction from the production of the field $\chi$ (recall the definition of $\varphi$ in eq. (\ref{varphi})). This relation is supplemented by the Friedman equation for the scale factor, where we also disregard the backreaction of $\chi$. Disregarding slow roll corrections, the potential $V \left( \phi \right)$ is constant, and we have the de-Sitter background solution $a = - \frac{1}{H \, \tau}$.

(ii) The inflation perturbations, in absence of their interaction with the quanta of $\chi$. We decompose them as 
\begin{eqnarray}
\delta \phi^{(0)} \left( \tau ,\, \vec{x} \right) &=&  \int \frac{d^3 k}{\left( 2 \pi \right)^{3/2}} \, {\rm e}^{i \vec{x} \cdot \vec{k}} \, \delta \phi^{(0)} \left( \tau ,\, \vec{k} \right) \;\;, \nonumber\\ 
\delta \phi^{(0)} \left( \tau ,\, \vec{k} \right) &=& \delta \phi_k^{(0)} \left( \tau \right) a_{\vec{k}} +  \delta \phi_k^{(0)*} \left( \tau \right) a_{-\vec{k}}^\dagger \;, 
\label{phi-deco}
\end{eqnarray} 
where the $\phi^{(0)}$ annihilation and creation operators obey the commutation relations $\left[  a_{\vec{k}} ,\, a_{\vec{k}'}^\dagger \right] = \delta^{(3)} \left( \vec{k} - \vec{k}' \right)$. From the unperturbed lagrangian, we obtain the mode function 
\begin{equation}
\delta \phi^{(0)}_k \left( \tau \right) =  \frac{H}{\sqrt{ 2 k}} \left( i \, \tau + \frac{1}{k} \right) {\rm e}^{- i k \tau} \;. 
\label{df0}
\end{equation} 
where we have imposed that the mode is in the standard adiabatic vacuum in the deep sub-horizon ($-k \tau \gg 1$) regime. From these relations, we obtain the unperturbed power spectrum in the super-horizon regime, 
\begin{equation}
P_\zeta^{(0)} \left( \tau ,\, k \right) = \frac{H^2}{\dot{\varphi}^2} \, \left( \frac{H}{2 \pi} \right)^2 \;\;,\;\; -k \, \tau \ll 1 \;\;. 
\label{P-0-res}
\end{equation}

To obtain (\ref{df0}) and (\ref{P-0-res}), we have also assumed de Sitter background evolution. This introduces an error in the spectral tilt 
of the power spectrum, $P_\zeta \propto k^{n_s-1}$. While the result (\ref{P-0-res}) is scale invariant, the correct spectral tilt is  $n_s -1 = 2 \eta - 6 \epsilon$, where $\epsilon$ and $\eta$ are the slow roll parameters 
\begin{equation}
\epsilon \equiv \frac{M_p^2}{2} \left( \frac{V'}{V} \right)^2 \;\;,\;\; 
\eta \equiv M_p^2 \;  \frac{V''}{V} \;. 
\label{epsilon-eta} 
\end{equation}
To obtain the correct spectral tilt, one must take into account slow roll corrections to the background evolution, as well as scalar metric perturbations (see \cite{Riotto:2002yw} for a detailed derivation).  We discuss this in Appendix \ref{app:dg}, where we include metric perturbations in all but the last term of ${\cal L}_0$. (Terms that originate from including metric perturbations in the last term of  ${\cal L}_0$ will instead be put in the interaction lagrangian, see eqs. (\ref{S0-gd}) and (\ref{Sint-dg}).) In the main text we disregard slow roll corrections, and we provide the dominant correction to the power spectrum $\delta P_\zeta$. Our final result (\ref{P-int-res})  is therefore understood as the  main signature from the particle production, up to ${\cal O } \left( \epsilon ,\, \eta \right) = {\cal O } \left( n_s - 1 \right) = 10^{-2}$ corrections.  A similar understanding applies to our calculation of the bispectrum.

(iii) The production of quanta of $\chi$, to zeroth order in their backreaction on the background dynamics. We discussed this computation extensively in \cite{Pearce:2016qtn}, so here we simply state the results, referring the interested reader to that work for details. We decompose the $\chi^{(0)}$ field analogously to eq. (\ref{phi-deco}), 
\begin{eqnarray}
 \chi^{(0)} \left( \tau ,\, \vec{x} \right) &=&  \int \frac{d^3 k}{\left( 2 \pi \right)^{3/2}} \, {\rm e}^{i \vec{x} \cdot \vec{k}} \,  \chi^{(0)} \left( \tau ,\, \vec{k} \right) \;\;, \nonumber\\ 
 \chi^{(0)} \left( \tau ,\, \vec{k} \right) &=&  \chi_k^{(0)} \left(\tau \right) b_{\vec{k}} +   \chi_k^{(0)*} \left( \tau \right) b_{-\vec{k}}^\dagger \;\;, 
\label{chi-deco}
\end{eqnarray} 
where the $\chi^{(0)}$ annihilation and creation operators obey the commutation relations $\left[  b_{\vec{k}} ,\, b_{\vec{k}'}^\dagger \right] = \delta^{(3)} \left( \vec{k} - \vec{k}' \right)$. We are interested in the corrections to the power spectrum of $\zeta$ from the $\chi$ particle production. This is a much greater effect than the loop quantum corrections from the $\chi$ field in absence of particle production. Therefore we disregagrd the latter effect, and set $\chi = 0$ at $\tau < \tau_*$. We then have 
\begin{equation}
\chi^{(0)}_k \left( t \right) = \frac{\theta \left( \tau - \tau_* \right)}{a \left( \tau \right)  \, \sqrt{2 \omega \left( \tau \right)} } 
 \left[ \alpha \left( k \right) \, {\rm e}^{-i \int_{\tau_*}^\tau d \tau' \omega \left( \tau' \right)} +  \beta \left( k \right) \, {\rm e}^{i \int_{\tau_*}^\tau d \tau' \omega \left( \tau' \right)} \right] \;, 
\label{dc0}
\end{equation} 
where the comoving frequency of the modes is given by 
\begin{equation}
\omega \left( \tau \right) = \sqrt{k^2 + a^2 \, g^2 \left( \varphi - \phi_* \right)^2} \simeq a \, g \, \vert  \varphi - \phi_* \vert \,, 
\end{equation}
and where the Bogolyubov coefficients read 
\begin{equation}
\alpha \left( k \right) = \sqrt{1+ {\rm e}^{-\pi \kappa^2}} \, {\rm e}^{i \, \alpha_k} \;\;,\;\; \beta \left( k \right) = {\rm e}^{-\frac{\pi}{2} \, \kappa^2} \,, 
\label{alpha-beta}
\end{equation}
with  
\begin{equation}
\kappa \equiv \frac{k}{a_*  \sqrt{g \; \vert \dot{\varphi}_* \vert}} \;\;\;,\;\;\; 
\alpha_k = {\rm Arg } \left[ \Gamma \left( \frac{1 + i \kappa^2}{2} \right) \right] + \frac{\kappa^2}{2} \left( 1 - \log \frac{\kappa^2}{2} \right) \,.
\label{kappa}
\end{equation} 
We have denoted with a subscript ${}_*$ quantities evaluated at $\tau=\tau_*$. In eq.~(\ref{alpha-beta}) we have taken the asymptotic late time value for the Bogolyubov coefficients. This is accurate a short time after $\tau_*$, when the adiabaticity condition $\omega' \ll \omega^2$ is satisfied again. One can verify that 
\begin{equation}
\tau \gg  \tau_* + \frac{ \delta t_* }{ a\left( \tau_* \right) } \;\;,\;\; \delta t_* \simeq \frac{1}{\sqrt{g \, \vert \dot{\varphi}_* \vert}} \;\; \Rightarrow \;\; \omega' \;,\; {\cal H } \, \omega ,\, k^2 \ll \omega^2 \,.
\label{adiabaticity} 
\end{equation} 
Namely, after this time, the frequency varies adiabatically, and the quanta of $\chi$ are non-relativistic. (In this estimate we have substituted the momentum of the quanta with the typical momentum $k_{\rm typical} = a_* \sqrt{g \; \vert \dot{\varphi}_* \vert}$ obtained from eq. (\ref{kappa}). As is already clear from (\ref{alpha-beta}), the number  of quanta of $\chi$, and the results of the computations presented below, are dominated by the quanta having such momentum.) Moreover, the time variation in the mode function due to the phase $\propto \omega$ is much greater than the variation due to the expansion of the Universe. 

In our computations we assume a quick violation of non-adiabaticity, namely $\delta t_* \ll \frac{1}{H}$. Therefore, the non-adiabatic phase only lasts for a very small fraction of an e-fold of inflation, and in our computations below we assume the expressions (\ref{alpha-beta}) to be valid already at $\tau = \tau_*^+$, with the exception of terms that logarithmically diverge at $\tau_*$. (For these terms only, we account for the fact that the expression (\ref{alpha-beta}) is valid only from $\tau_* + \delta \tau_*$.) This approximation is very accurate, since, apart from these terms, the time integrands that we evaluate are regular at $\tau_*$, and so any error is suppressed by the smallness of $\delta t_*$.

In this approximation, the mode functions $\chi^{(0)}$ are discontinuous at $\tau_*$. This introduces $\delta \left( \tau_* \right)$ contributions that should be accounted for in the in-in computation of $\delta \phi$. These effects have been evaluated with different methods in~\cite{Elgaroy:2003hp,Romano:2008rr}, where it has been found that they lead to a change of the power spectrum that is parametrically smaller than the one that we obtain below. Therefore, we disregard such terms in our computation. 

The mode functions of $\chi$ modify  the power spectrum of $\zeta$ through $2-$point expectation values $\langle \chi^2 \rangle$. We renormalize them following the adiabatic renormalization procedure as we also did in  \cite{Pearce:2016qtn}, accounting only for the effect of particle production, 
\begin{eqnarray} 
&&\left\langle  :  \chi^{(0)} \left( \tau ,\, \vec{k} \right)  \chi^{(0)} \left( \tau' ,\, \vec{p} \right) : \right\rangle'  = 
\frac{\theta \left( \tau - \tau_* \right) \theta \left( \tau' - \tau_* \right) \, }{2 
a \left( \tau \right) a \left( \tau' \right) \, \sqrt{\omega \left( \tau \right) \omega \left( \tau' \right) }} \nonumber\\
&&\qquad\qquad\qquad\qquad\times
 \left[ \vert \beta_k \vert^2 \Phi \left( \tau \right)  \Phi^* \left( \tau' \right) 
+ \alpha_k \beta_k^* \Phi^* \left( \tau \right)  \Phi^* \left( \tau' \right) + {\rm c.c.} \right]\,, \label{:chi-chi:}
\end{eqnarray} 
where $\Phi \left( \tau \right) \equiv {\rm e}^{i \int_{\tau_*}^\tau \, d {\tilde \tau} \, \omega \left( {\tilde \tau} \right)}$.
%

We use this expression to evaluate the energy density of the produced $\chi-$quanta, 
\begin{equation} 
\rho_{\chi,{\rm bck}} = \left\langle \rho_\chi^{(0)} \right\rangle = \frac{1}{2 a^2} \left\langle : \chi^{(0)'} {}^2 + \partial_i \chi^{(0)}  \partial_i \chi^{(0)} + a^2 g^2 \left( \varphi - \phi_* \right)^2 \chi^{(0)2} : \right\rangle \,. 
\end{equation} 
From the decomposition (\ref{chi-deco}) and  the regularized expression (\ref{:chi-chi:}) we obtain 
\begin{equation}
\rho_{\chi,{\rm bck}} \simeq \frac{\theta \left( \tau - \tau_* \right)}{a^4 \left( \tau \right) } \int \frac{d^3 p}{\left( 2 \pi \right)^3} \, \omega \left( \tau ,\, p \right) \, \left\vert \beta_p \right\vert^2 \;, 
\end{equation} 
where the conditions  (\ref{adiabaticity}) have been used in the time differentiation. As discussed above, $\omega \simeq a \, g \, \vert \varphi - \phi_* \vert$ in the regime where the number density of $\chi-$quanta is non negligible. This gives 
\begin{equation}
\rho_{\chi,{\rm bck}} \simeq  \theta \left( \tau - \tau_* \right)    \frac{ g \,  \vert \varphi - \phi_* \vert \,  \vert g \, \dot{\varphi}_* \vert^{3/2}}{8 \pi^3} \, 
 \left( \frac{a_*}{a} \right)^3  \;. 
\label{rho-chi-bck}  
\end{equation} 
Namely, after they are produced, the quanta of $\chi$ are massive particles with an adiabatically varying mass. Disregarding the slow time variation of the mass, their energy density redshifts as that of matter, thus justifying our statement above.

\subsection{Corrections to the power spectrum due to particle production  } 
\label{subsec:int}

We split the interaction hamiltonian following from (\ref{action}) into two terms 
\begin{eqnarray} 
H_{\rm int}^{(0)} \left( \tau \right) &=& H_{{\rm int},1}^{(0)} \left( \tau \right) +  H_{{\rm int},2}^{(0)} \left( \tau \right) \;\;, \nonumber\\ 
H_{{\rm int},1}^{(0)} \left( \tau \right) &=&  a^4 \left( \tau \right)  \,  g^2 \left[ \varphi \left( \tau \right)  - \phi_* \right] \, \int d^3 x \, 
 \delta \phi^{(0)} \left( \tau ,\, \vec{x} \right)  \, \chi^{(0)2} \left( \tau ,\, \vec{x} \right)  \;\;, \nonumber\\ 
H_{{\rm int},2}^{(0)} \left( \tau \right) &=&  a^4 \left( \tau \right) \,  \frac{g^2}{2}  \, \int d^3 x  \, \delta \phi^{(0) 2} \left( \tau ,\, \vec{x} \right)  \, \chi^{(0)2} \left( \tau ,\, \vec{x} \right)  \;\;. 
\label{Hint-12}
\end{eqnarray}

\begin{figure}[ht!]
\centerline{
\includegraphics[width=0.8\textwidth,angle=0]{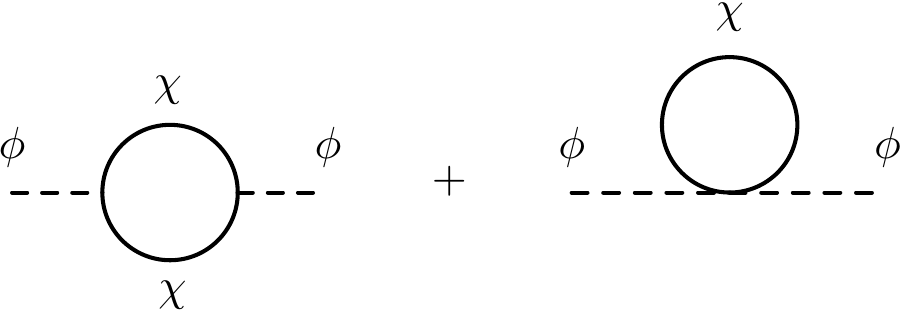}
}
\caption{
Diagrammatic expressions  that contribute to the corrections 
$\delta_1 P_\zeta$ and $\delta_2 P_\zeta$ 
written in eqs. (\ref{d1P}) and (\ref{d2P}). 
}
\label{fig:diagrams}
\end{figure}

Once inserted into eq. (\ref{P-int}), these two terms provide the one loop corrections 
\begin{equation}
\delta_1 P_\zeta \left( \tau ,\, k \right) = - \frac{k^3}{2 \pi^2} \, \frac{\varphi^{' 2} \left( \tau \right)}{{\cal H}^2 \left( \tau \right)} \, 
\int_{\tau_*}^{\tau} d \tau_1 \, \int_{\tau_*}^{\tau_1} d \tau_2 \, 
\left\langle \left[ \left[  \delta \phi^{(0)} \left( \tau ,\, \vec{k} \right)  \delta \phi^{(0)} \left( \tau ,\, \vec{k}' \right)  ,\, 
H_{{\rm int},1}^{(0)} \left( \tau_1 \right) \right] ,\, H_{{\rm int},1}^{(0)} \left( \tau_2 \right) \right] \right\rangle ' \,, 
\label{d1P}
\end{equation} 
and 
\begin{equation}
\delta_2 P_\zeta \left( \tau ,\, k \right) = - i \, \frac{k^3}{2 \pi^2} \, \frac{\varphi^{' 2} \left( \tau \right)}{{\cal H}^2 \left( \tau \right)} \, 
\int_{\tau_*}^{\tau} d \tau_1
\left\langle \left[  \delta \phi^{(0)} \left( \tau ,\, \vec{k} \right)  \delta \phi^{(0)} \left( \tau ,\, \vec{k}' \right)  \,, 
H_{{\rm int},2}^{(0)} \left( \tau_1 \right) \right] \right\rangle ' \;,  
\label{d2P}
\end{equation} 
diagrammatically shown in Figure \ref{fig:diagrams}. The cubic vertices in the first diagram are proportional to the inflaton homogeneous value, and this diagram accounts for the rescattering of $\chi$ quanta off the inflaton condensate, producing inflaton perturbations. The right diagram instead originates from the quartic coupling of  two inflaton and two $\chi$ quanta.

\begin{figure}
\centerline{
\includegraphics[width=0.4\textwidth,angle=0]{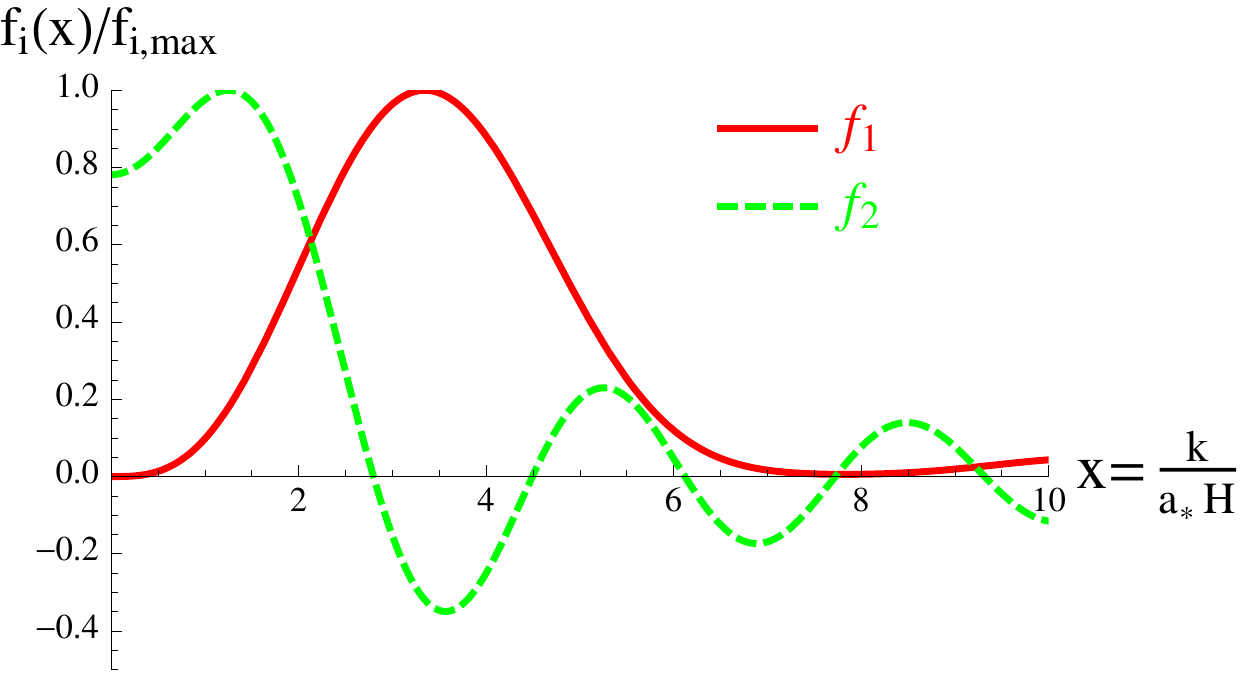}
\includegraphics[width=0.4\textwidth,angle=0]{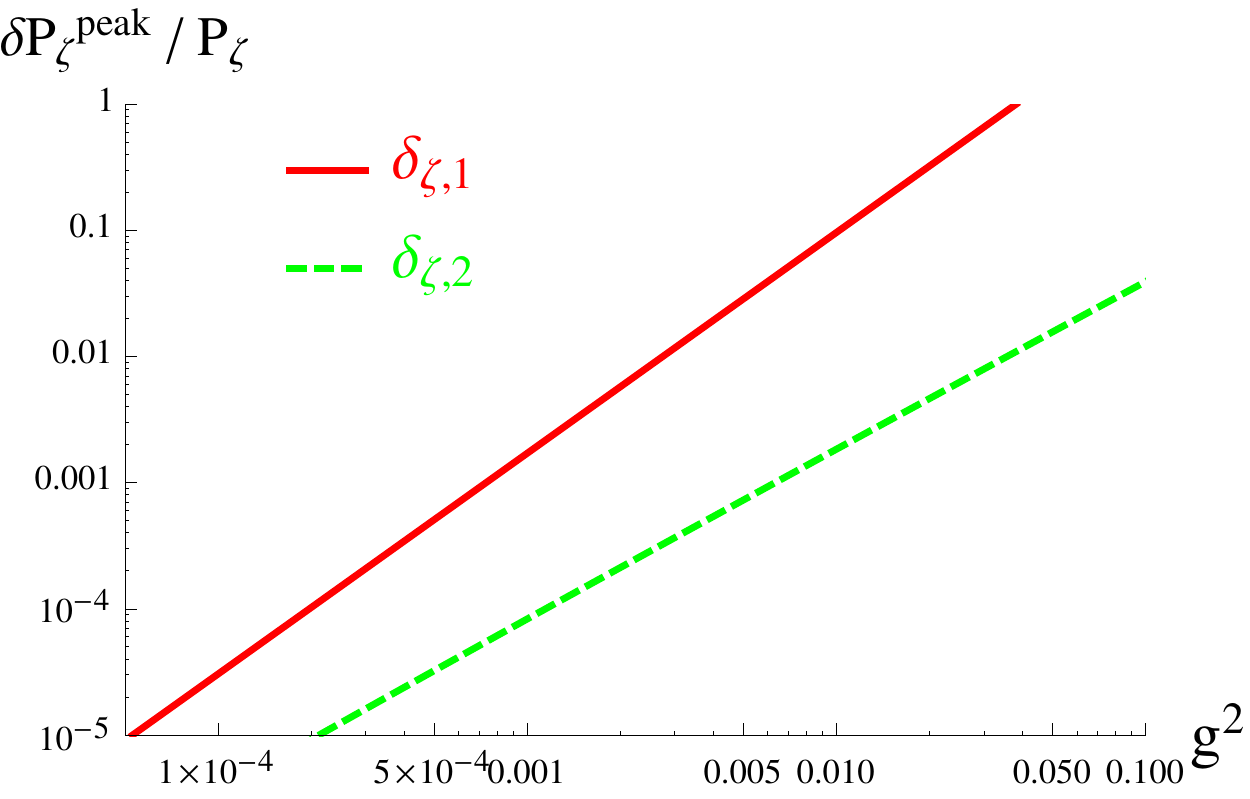}
}
\caption{Left panel: shape of the corrections, normalized to one at its maximum. The actual value of the maxima are $f_{1,{\rm max}} \simeq 0.11$ and  $f_{2,{\rm max}} \simeq 0.85$. Right panel: relative magnitude of the dominant correction $\delta_{\zeta,2}$ at their maximum, vs. the standard PS. The correction amounts to $1\%$ (resp. $10\%$) for $g^2 \simeq 0.0027$ (resp. $g^2 \simeq 0.01$).  
}
\label{fig:corrections2}
\end{figure}

We evaluate these contributions in Appendix \ref{app:PS}. We find 
\begin{eqnarray}
\delta_{\zeta,1} \left( k \right) &\equiv& \frac{\delta_1 P_\zeta \left( k \right)}{ P_\zeta^{(0)} \left( k \right)} \simeq  \frac{\left(2+\sqrt{2}\right) \; g^2 \; \left(  g \vert \dot{\varphi}_* \vert \right)^{3/2} }{8 \, \pi^3 \; H^3}    \; f_1 \left( \frac{k}{a_* \, H} \right) \;, \nonumber\\ 
\delta_{\zeta,2} \left( k \right) &\equiv& \frac{\delta_2 P_\zeta \left( k \right)}{ P_\zeta^{(0)} \left( k \right) } \simeq  \frac{ g^2 \, \sqrt{g \, \vert \dot{\varphi}_* \vert } }{8 \pi^3 H}   \;  \ln \left(  \frac{\sqrt{g \, \vert \dot{\varphi} \vert_*}}{H} \right)  \; f_2 \left(  \frac{k}{a_* \, H}  \right)  \;,   
\label{P-int-res} 
\end{eqnarray} 
where $f_{1,2} \left( x \right)$ are two dimensionless functions that govern the scale dependence of the corrections 
\begin{eqnarray}
f_1 \left( x \right) \equiv   \frac{ \left[ \sin \left( x \right) - {\rm SinIntegral} \left( x \right) \right]^2 }{x^3}  \;\;,\;\;  
f_2 \left( x \right) \equiv \frac{ -2 x \,  \cos \left( 2 x \right) + \left( 1 - x^2 \right)  \sin \left( 2 x \right) }{x^3}  \,, \nonumber\\ 
\label{shapes}
\end{eqnarray} 
which we show in the left panel of Figure \ref{fig:corrections2}.

The spectral shape of the corrections has a peak at $k = {\rm O } \left( a_* \, H \right)$, namely at the scale parametrically given by the momentum of the mode that left the horizon while particle production occurred, followed by smaller oscillations. More precisely, the peak of the function $f_1$ occurs at  $x \simeq 3.35$, while the function evaluates to $f_1 \simeq 0.11$. The peak of the function $f_2$ occurs at  $x \simeq 1.25$, while the function evaluates to $f_2 \simeq 0.85$. 

Under the assumption that the corrections are subdominant, $\vert \delta_{\zeta,i} \vert \ll 1 $, we impose that the zeroth order power spectrum (\ref{P-0-res}) matches the observed value  $P_\zeta \simeq 2.2 \cdot 10^{-9} $  \cite{Ade:2015lrj}, which implies $\sqrt{\vert \dot{\varphi} \vert} / H \simeq 58$. Therefore, the values of the two corrections at their peak are 
\begin{eqnarray}
\delta_{\zeta,1}  \vert_{\rm peak} & \simeq & 0.0015 \, g^2 \,  \left( \frac{\sqrt{g \, \vert \dot{\varphi} \vert}}{H} \right)^3  
\simeq 300 \, g^{7/2}  \;\;\;\;\;\; {\rm at} \;\;\;\;\;\; \frac{k}{a_* \, H} \simeq 3.35  \;\;,   \nonumber\\  
\delta_{\zeta,2}  \vert_{\rm peak} & \simeq & 0.0034 \, g^2 \,  \frac{\sqrt{g \, \vert \dot{\varphi} \vert}}{H} \,  \ln \left(  \frac{\sqrt{g \, \vert \dot{\varphi} \vert}}{H} \right)   \simeq 0.05 \, g^{5/2} \,{\rm ln } \left( \frac{g}{0.0003} \right)^2    \;\;\;\;\;\; {\rm at} \;\;\;\;\;\; \frac{k}{a_* \, H} \simeq 1.25  \,. \nonumber\\ 
\label{dPZ-peak} 
\end{eqnarray} 
In the right panel of Figure \ref{fig:corrections2} we show the peak values of $\delta_{\zeta,1}$ and of  $\delta_{\zeta,2}$ as a function of the coupling $g^2$. We see that the correction $\delta_{\zeta,1}$ dominates. We find that the correction amounts to $1\%$ (resp. $10\%$) for $g^2 \simeq 0.0027$ (resp. $g^2\simeq 0.01$). These values of $g^2$ are well within the regime of validity of perturbation theory, see eq. (\ref{perturbativity}).

\subsection{Contribution to the bispectrum from particle production } 
\label{subsec:BS}

\begin{figure}[ht!]
\centerline{
\includegraphics[width=0.8\textwidth,angle=0]{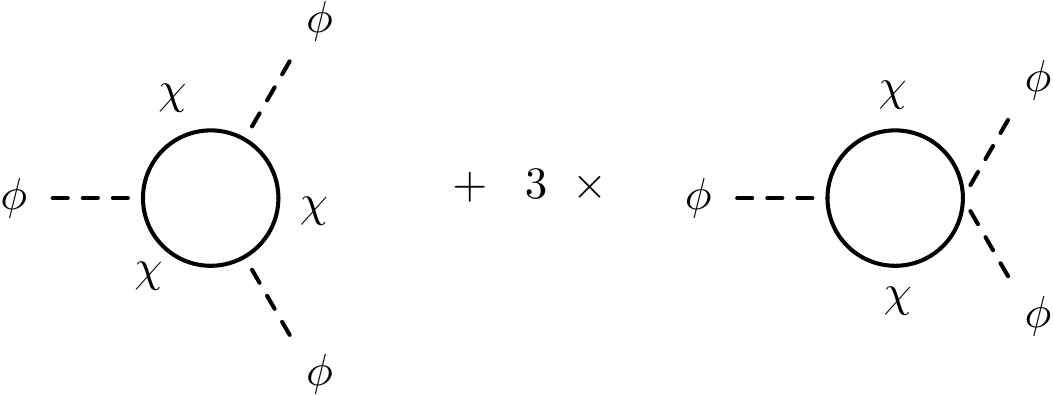}
}
\caption{
Diagrammatic expressions  that contribute to the bispectrum from particle production. The $\times 3$ denotes the three permutations of the second diagram over the three vertices.}
\label{fig:diagramsbS}
\end{figure}

In this subsection we compute the contribution to the bispectrum 
\begin{equation}
B \left( \tau ;\, k_1 ,\, k_2 ,\, k_3 \right) \equiv \left\langle \zeta \left( \tau ,\, \vec{k}_1 \right)  \zeta \left( \tau ,\, \vec{k}_2 \right)  \zeta \left( \tau ,\, \vec{k}_3 \right) \right\rangle' \;, 
\end{equation} 
from particle production. (We recall that the prime denotes an expectation value without the corresponding $\delta^{(3)} \left( \sum_i \vec{k}_i \right)$ function). 
We disregard the zeroth order bispectrum (that is, the one in the absence of particle production), which is known to be negligible. We instead compute the dominant contribution from particle production through the in-in formalism, completely analogous to the computation of the power spectrum in the previous subsection. Specifically, we compute the terms that are diagrammatically represented in Figure \ref{fig:diagramsbS}. We discuss the details of the computation in Appendix \ref{app:BS}.  Analogously to what happens for the power spectrum, the diagram with only trilinear vertices  (namely, the  left diagram in Figure \ref{fig:diagramsbS}) dominates the bispectrum, leading to 
\begin{eqnarray} 
B \left( \tau ;\, k_1 ,\, k_2 ,\, k_3 \right)  \Big\vert_{k_1=k_2=k_3 \equiv k} 
&&  \simeq  \frac{1}{k^6} \frac{H^3}{\vert \dot{\varphi}_* \vert^{3/2}}  \, g^{9/2} \,  \frac{\left( 27 + 8 \, \sqrt{6} \right)}{288 \pi^{9/2}} \,  
f_3 \left( \frac{k}{a_* \, H} \right)  \,, \nonumber\\ 
\label{eq:BS}
\end{eqnarray} 
where we have introduced the new shape function 
\begin{equation}
f_3 \left( x \right) \equiv \left( \frac{  {\rm SinIntegral } \left( x \right)  -    \sin \left( x \right)  }{x}  \right)^3 \,. 
\end{equation} 

\begin{figure}
\centerline{
\includegraphics[width=0.4\textwidth,angle=0]{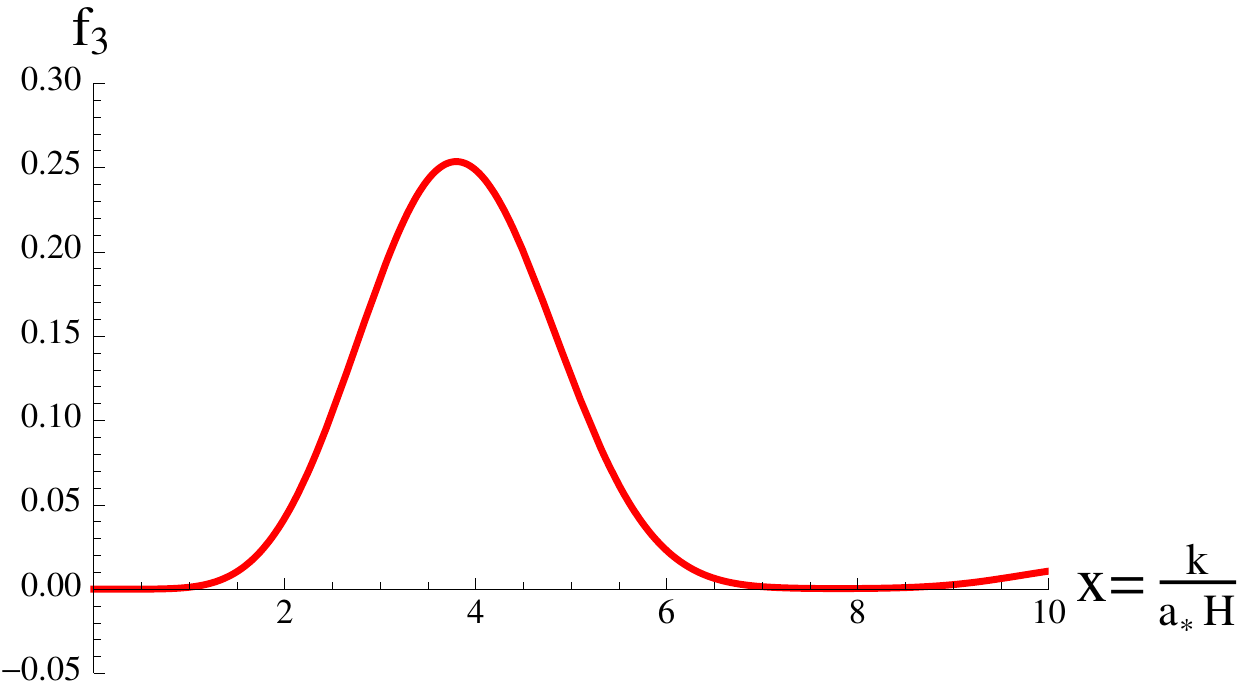}
\includegraphics[width=0.4\textwidth,angle=0]{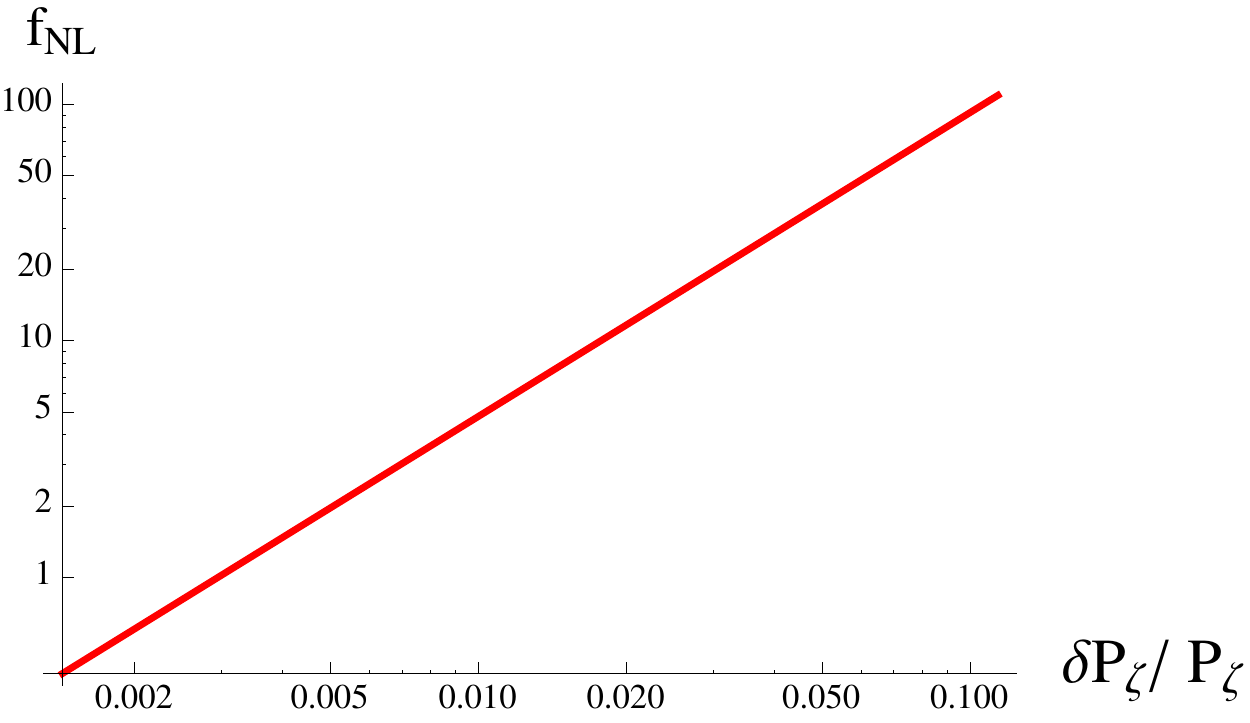}
}
\caption{Left panel: scale dependence of the bispectrum (as defined in eq. (\ref{fNL-peak})), evaluated on an exactly equilateral configuration.   Right panel: Peak value of $f_{\rm NL} \left( x,\, x ,\, x \right)$ (eq. (\ref{fNL-peak})) vs. the peak value of the correction of the power spectrum, eq.  (\ref{dPZ-peak}). 
}
\label{fig:corrections3}
\end{figure}

It is conventional to introduce the nonlinearity parameter 
\begin{equation}
B_\zeta \left( k_1 ,\, k_2 ,\, k_3 \right) = \frac{3}{10} \left( 2 \pi \right)^{5/2} \, 
f_{NL} \left( k_1 ,\, k_2 ,\, k_3 \right) \, P_\zeta^2  \, \frac{\sum_i k_i^3}{\prod_i k_i^3} \,,  
\label{fNL-def}
\end{equation} 
where the numerical coefficient follows from the $2 \pi$ convention adopted here, see e.g.~\cite{Barnaby:2011vw}. This gives 
\begin{equation}
f_{\rm NL} \left( k ,\, k ,\, k \right) \simeq 10^7 \, g^{9/2} \, f_3 \left( \frac{k}{a_* H} \right) \,, 
\end{equation}
where we have used the unperturbed value for the power spectrum $P_\zeta \simeq P_\zeta^{(0)} \simeq 2.2 \cdot 10^{-9} $  \cite{Ade:2015lrj} 
(which, as we have already remarked, gives  $\sqrt{\vert \dot{\varphi} \vert} / H \simeq 58$). As seen from the left panel of Figure \ref{fig:corrections3}, the bispectrum also exhibits a bump at a scale parametrically close to that of the modes that left the horizon during the episode of particle production, at which the sourced power spectrum is peaked as well. The peak of the function $f_3$ occurs at $x \simeq 3.8$, where the function evaluates to $f_3 \simeq 0.25$. This gives 
\begin{equation}
f_{\rm NL} \left( k ,\, k ,\, k \right)  \vert_{\rm peak}  \simeq \left( \frac{g^2}{0.0014} \right)^{9/4} \;\;\;\; {\rm at } \;\;\;\; k \simeq 3.8 \, a_* \, H \;. 
\label{fNL-peak}
\end{equation}

We recall that in the previous subsection we found that, for $g^2 \simeq 0.0027$, the peak value of the sourced power spectrum is about $1\%$ of the vacuum power spectrum. For this value of $g^2$ we find  $f_{\rm NL} \left( k ,\, k ,\, k \right)  \vert_{\rm peak} \simeq 4.6$. A $10\%$ correction to the power spectrum is instead obtained for  $g^2 \simeq 0.01$. For this value we find  $f_{\rm NL} \left( k ,\, k ,\, k \right)  \vert_{\rm peak} \simeq 88$. In the right panel of Figure \ref{fig:corrections3} we show the peak value of $f_{\rm NL}$ vs. the peak value of the correction of the power spectrum. 

\section{Localized particle production in a hidden sector}
\label{sec:hidden}

In this section we study the imprint on the scalar power spectrum of a sudden episode of  production of $\chi-$particles  due to a field $\psi$ which is different  from the inflaton. We assume that this production occurs in a sector that is only gravitationally coupled to the inflaton field; namely the action of the model is 
\begin{equation}
S = \int d^4 x \sqrt{-g} \left[  \frac{M_p^2}{2} \, R - \frac{1}{2} \left( \partial \phi \right)^2 - V_\phi \left( \phi \right)  - \frac{1}{2} \left( \partial \psi \right)^2 
 - \frac{1}{2} \left( \partial \chi \right)^2 - V_\psi \left( \psi \right) - \frac{g^2}{2} \left( \psi - \psi_* \right)^2 \chi^2 \right]  \;. 
\label{3fld}
\end{equation} 

We assume that the homogeneous part of the field $\psi$ crosses the value $\psi_*$ during inflation, at a time in which the CMB or LSS modes exited the horizon. We also assume that at this moment the field $\psi$ is in slow roll,  defining a slow roll parameter $\epsilon_\psi$ through the relation $\dot{\psi}^2 = 2 \epsilon_\psi \,  H^2 \, M_p^2$, and that the field $\psi$ continues this motion, with a nearly constant velocity (apart from the small correction due to the backreaction of the produced $\chi$ particles) and small $\epsilon_\psi$, for $N$ e-folds after the instance of particle production. Finally, we assume that  the inflaton $\phi$ dominates the energy density of the universe during inflation, and that the inflaton background evolution is always in the slow roll regime, with $\epsilon_\phi \ll 1$ and $\dot{\phi}^2 \simeq 2 \epsilon_\phi^2  \, H^2 \, M_p^2$ either constant or adiabatically evolving. 

The production of quanta of $\chi$, and their subsequent evolution, proceed exactly as described in Section \ref{subsec:zero}, and the relations (\ref{alpha-beta}) for the occupation number continue to hold, with $\dot{\varphi}_*$ replaced by  $\dot{\psi}_*$. After they are produced, the quanta of $\chi$ are massive particles with a nearly constant mass, and therefore they are diluted  as matter  by the inflationary expansion. Before being diluted, the quanta can source $\psi$ perturbations through the direct coupling between the $\chi$ and $\psi$ fields in the potential of (\ref{3fld}). They can also source  inflaton perturbations gravitationally (through interactions that are technically obtained by integrating out metric perturbations).  The perturbations of $\psi$ sourced by the $\chi-$quanta act in their turn as a source of inflation perturbations, through a quadratic $\delta \phi \, \delta \psi$ mixing that exists as long as the both fields are evolving (see below). 

Generally speaking, well after particle production, when the $\chi-$quanta have been diluted away, the scalar curvature is a linear combination $\zeta \left( \tau ,\, \vec{k} \right) = c_1 \left( \tau \right) \delta \phi \left( \tau ,\, \vec{k} \right) +  c_2 \left( \tau \right) \delta \psi \left( \tau ,\, \vec{k} \right)$, where the two coefficients $c_1$ and $c_2$ depend on the background evolution of the two fields.  (The orthogonal combination is an isocurvature mode.) We assume that the field $\psi$ rolls for a finite number of e-folds $N < 60$ after the particle production, and that it gets stabilized at some value $\psi_0$ well before the end of inflation. In this way its energy (both at the background and at the perturbative level) becomes negligible with respect to that of the inflaton, so that in the final stages of inflation and at reheating, the inflaton perturbations highly dominate the curvature perturbation ($c_2 / c_1 \rightarrow 0$). Therefore, the standard relation  $\zeta \simeq - \frac{H}{\dot{\varphi}} \, \delta \phi$ continues to provide an accurate approximation for the scalar curvature perturbations at the end of inflation and at reheating. In this framework, the perturbations of the $\chi$ and $\psi$ fields affect observables only to the degree to which they affect the perturbations of the inflaton field though the couplings described in the previous paragraph. 

The relevant interactions that contribute to this observable are 
\begin{align}
 H_\mathrm{int}(\tau)
&\supset \int d^3x \, a^4 \Bigg\{ g^2 \left( \psi - \psi_* \right) \delta \psi \, \chi^2 + \frac{g^2}{2} \delta \psi^2 \, \chi^2 
- 6 \, {\rm sign } \left( \dot{\varphi} \dot{\psi} \right) \sqrt{\epsilon_\phi \epsilon_\psi} H^2 \delta \phi \, \delta \psi 
 \nonumber\\ 
&  \qquad  \qquad  \qquad 
 + \frac{\mathrm{sgn}(\dot{\varphi})}{2 a^2} \sqrt{ \dfrac{\epsilon}{2} } \dfrac{\delta \phi}{M_p} \left(\chi^\prime \chi^\prime + \partial_i \chi \partial_i \chi + g^2 a^2 (\varphi - \phi_*)^2 \chi^2 \right)  \nonumber \\
&  \qquad  \qquad  \qquad  - \frac{ {\rm sgn} ( \dot{\varphi} ) H}{a} \sqrt{\frac{\epsilon}{2}}  \chi' \, \partial_i \chi  \; \Delta^{-1} \left[ \frac{\partial_i \delta \phi^{'} }{a \,  H \, M_p} + \left( \eta - 2 \epsilon \right)    \, \frac{\partial_i \delta \phi }{M_p} \right]  \Bigg\} \,.
\label{Hint-3fld}
\end{align}
The first two terms are obtained from the last term in (\ref{3fld}). They are analogous to the interaction term in (\ref{action}), with the difference that now $\chi$ is directly coupled to $\psi$, and not to the inflaton. The third term originates from the linearized theory of cosmological perturbations with two scalar fields. 
It is the extension to two fields of eq. (\ref{S2-metric}). The second and third lines are gravitational interactions obtained by integrating out the scalar metric perturbations $\delta g_{00}$ and $\delta g_{0i}$ (analogously to the second and third line of eq. (\ref{eq:Hint_v1})).

\begin{figure}
\centerline{
\includegraphics[width=\textwidth,angle=0]{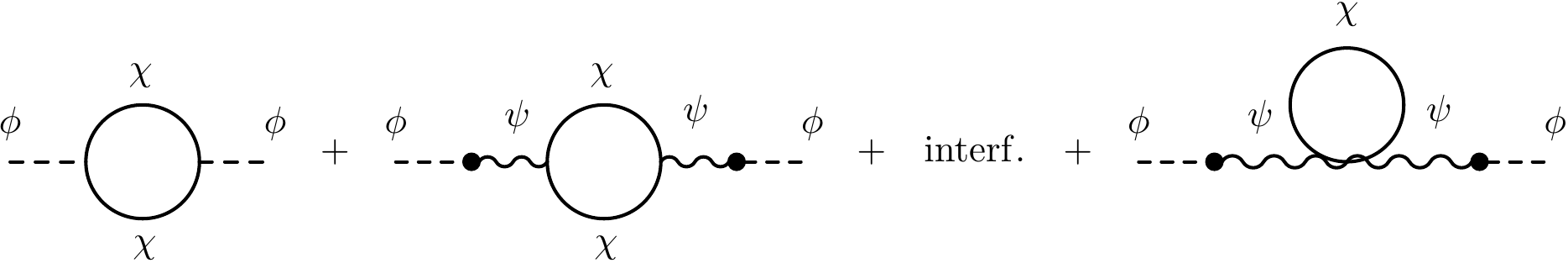}
}
\caption{One diagrams contributing $\delta P_\zeta$ in the model (\ref{3fld}). 
}
\label{fig:3fld-diag}
\end{figure}

In Figure \ref{fig:3fld-diag} we show the one loop diagrams obtained from (\ref{Hint-3fld})  that account for the effects of particle production on $P_\zeta$. 
The first diagram shown in the figure has two gravitational $\chi^2 \delta \phi$ vertices, each obtained from the last two lines of (\ref{Hint-3fld}). The second diagram is obtained from two direct cubic $\chi^2 \delta \psi$ interactions. Each mode $\delta \psi$ is then ``converted'' into an inflaton perturbation through 
a mass insertion, originating from the third term in (\ref{Hint-3fld}). We do not explicitly show the two diagrams that give the interference between the first two diagrams in the figure, namely processes that have one $\chi^2 \delta \phi$ interaction on one side of the $\chi$ loop, and the  $\chi^2 \delta \psi$ + $\delta \psi \delta \phi$ interactions on the other side. The final diagram shown in the figure is obtained from the direct quartic $\chi^2 \delta \psi^2$ interaction and two mass insertions. 

In comparing the second and the last diagram shown in Figure  \ref{fig:3fld-diag} we recall the hierarchy between the two diagrams shown in Figure \ref{fig:corrections2}. We found there that the diagram with two cubic vertices is parametrically dominant over the one with the quartic interaction 
(see also the right panel of Figure \ref{fig:corrections2}). This hierarchy can be understood from the Feynman rules that we work out in Appendix \ref{app:Feynman}. The same hierarchy exists between the second and the last diagram of Figure \ref{fig:3fld-diag}, so we can disregard the last diagram. 

For the effect of the first diagram, we can use the result obtained in Appendix \ref{app:dg}, eqs. (\ref{dP2-grav-full}) and (\ref{dP2-grav}). Those equations refer to the same diagram, with the only difference that in the case studied in Appendix  \ref{app:dg} the $\chi-$quanta were produced by the inflaton field. Therefore the expression $\vert \dot{\varphi}_* \vert^{3/2}$ appearing in those expressions (which is related to the number density of the $\chi-$quanta) must be replaced by  $\vert \dot{\psi}_* \vert^{3/2}$. Additionally, one factor of $\epsilon$ in those results is related to the derivative of the field responsible for particle production (technically, as expressed in eq. (\ref{mass-app})), so it should be $\epsilon_\psi$ in this case. The final factor of $\epsilon$ in these two formulae is related to the gravitational coupling between $\chi^2$ and $\delta \phi$, which in this case is written as $\epsilon_\phi$. Taking this into account, the first diagram in Figure \ref{fig:3fld-diag} gives a localized correction to the scalar power spectrum, which has the maximum value 
\begin{eqnarray} 
\frac{\delta P_{\phi,{\rm first \; diagram}}}{P_\phi^{(0)}} \Big\vert_{\rm peak} & \simeq & 4 \cdot 10^{-4} \,  \frac{  g^{7/2} \; \vert \dot{\psi}_* \vert^{3/2} }{H^3} \, \epsilon_\phi \, \epsilon_\psi  \;\;\; {\rm at } \;\;\; \frac{k}{a_* \, H} \simeq 4.67 \;. 
\label{3fld-one}
\end{eqnarray} 

The second diagram is instead evaluated in Appendix \ref{app:3fld-dominant}, and it also gives a localized correction  to the scalar power spectrum, which has the maximum value (see eq. (\ref{3fld-dominant-res}) for the full shape) 
\begin{equation}
\frac{\delta P_{\phi,{\rm second \; diagram}}}{P_\phi^{(0)}} \Big\vert_{\rm peak}  \simeq 6 \cdot 10^{-3}   \, \frac{  g^{7/2} \, \vert \dot{\psi}_* \vert^{3/2} }{ H^3 }  \,  \epsilon_\phi \, \epsilon_\psi \, N_k^2 \;\;\; {\rm at } \;\;\; \frac{k}{a_* \, H} \simeq 3.35 \;, 
\label{3fld-two}
\end{equation} 
where $N_k$ is the number of e-folds between the moment the mode leaves the horizon and the moment that $\psi$ stops rolling. (It arises because the mode $\delta \psi$, produced by its direct coupling to the $\chi-$quanta, continues to source inflaton perturbations through the quadratic coupling, which is effective as long as both $\varphi$ and $\psi$ are  rolling.) The result (\ref{3fld-two}) is numerically enhanced with respect to (\ref{3fld-one}) already for  $N_k = {\rm O } \left( 1 \right)$, and it is further parametrically enhanced by $N_k^2$. Therefore the diagram with the two mass insertions dominates the sum, including the two interference terms, of the processes outlined in Figure \ref{fig:3fld-diag}.

\section{Discussion}
\label{sec:discussion}

In this work we have studied the imprint  on the power spectrum and bispectrum of the scalar curvature $\zeta$ due to a localized episode of particle production during inflation. For definiteness we have considered a specific mechanism \cite{Chung:1999ve}, described by the action  (\ref{action}),  in which a field $\chi$ is coupled to the inflaton $\phi$ through a $g^2 \left( \phi - \phi_* \right)^2 \chi^2 / 2$ interaction. As the inflaton moves past $\phi_*$, a burst of quanta of $\chi$ is produced. These quanta backreact on the background evolution of the inflaton and of the scale factor, and they source inflaton perturbations through the same interaction term that produced them. 

After they are produced, the quanta of $\chi$ are rapidly diluted by the inflationary expansion. They therefore provide a negligible direct contribution to the scalar curvature a few e-folds after the production; nor do they provide any isocurvature mode. We have verified this explicitly in Appendix \ref{app:zeta}, where  we have shown that the standard relation $\zeta \simeq - \frac{H}{\dot{\varphi}} \, \delta \phi$, typical of single field inflation, remains a very accurate approximation at large scales and a few e-folds after the particle production. As a further simplification,   in Appendix  \ref{app:dg}  we have demonstrated  that the contribution of the metric perturbations to the sourced part of $\zeta$ can also be disregarded. Specifically, we have proved there that the direct contribution of the metric perturbations to $\zeta$ is negligible, and that  the additional interactions between the $\chi$ and $\phi$ quanta mediated by the metric perturbations are suppressed with respect to the direct interactions from the coupling term in the potential.~\footnote{We work in the spatially flat gauge, where the  only metric perturbations are the nondynamical modes $\delta g_{00}$ and $\delta g_{0i}$. These modes are integrated out and induce additional interactions between the $\chi$ and the inflaton quanta. }

We performed computations in the in-in formalism, using as the interaction hamiltonian the couplings between the $\chi$ quanta and the inflaton perturbations. At the unperturbed level, this scheme gives the standard slow roll motion of the background inflaton  and the production of quanta of $\chi$ from this motion. From the interaction hamiltonian, we obtained the leading correction to the background inflaton motion,~\footnote{Specifically, in Appendix  \ref{app:dzeta} we showed how a tadpole diagram accounts for this backreaction in Hartree approximation.}  as well as on the perturbations.  We (for the first time, to our knowledge) developed Feynman rules for this model, described in Appendix  \ref{app:Feynman}, which provide correct order of magnitude estimates for any diagram 
that contributes to the $\zeta$ correlators, and that  allow us to quantify for which values of the coupling constant $g^2$ the computations are under perturbative control. 

We confirm the findings of~\cite{Barnaby:2009mc,Barnaby:2009dd,Barnaby:2010ke} that the dominant effect on the power spectrum is due to the inflaton perturbations produced (at the non-linear level) by rescattering of quanta of $\chi$ on the inflaton condensate.  This production generates a bump in the power spectrum for modes that left the horizon close to the moment at which the sudden particle production took place. As discussed in the Introduction, we improve over these works  by providing the precise analytical shape and height of the bump, as a function of the coupling constant $g^2$ and of the ratio $\sqrt{\vert \dot{\varphi} \vert} / H$ between the square root of the derivative of the inflaton and the Hubble rate.  The bump in the sourced modes needs to be subdominant with respect to the standard vacuum modes. Imposing that  the amplitude of the vacuum modes is properly normalized  \cite{Ade:2015lrj} gives $\sqrt{\vert \dot{\varphi} \vert} / H \simeq 58$, so that the shape and height of the bump only depend on $g^2$. We find that the ratio between the sourced and vacuum power spectrum at the location of the bump is given by  $300 \, g^{7/2}$.  

The correction from the peak can be as large as $10\%$ for $g^2 \approx 0.01$, and remains above $1\%$ down to $g^2 \gtrsim 0.003$.  If $k \sim a_* H$, the scale leaving the horizon during particle production,  is $\sim 10^{-5} \, \mathrm{Mpc}^{-1}$, then the peak can be probed by CMB observations; at larger $k$, constraints from weak lensing and large scale structure may become relevant.  Specific constraints on features in the spectrum have been explored in the literature; for example, Ref.~\cite{Barnaby:2009dd} employed a Markov chain Monte Carlo calculation using large scale structure and CMB data to constrain the height of the peak to be less than $10\%$.   Subsequent to this, the Planck collaboration has studied potential spectral features, most recently in~\cite{Ade:2015lrj}. Section 9 in that paper analyzes a number of models that lead to features. In particular, they study the form of the bump in the power spectrum that emerges from a step in the inflationary potential \cite{Miranda:2013wxa}. The best fit amplitude for such a bump is of about 35\% of the amplitude of the vacuum fluctuations -- see table 13. The authors of~\cite{Ade:2015lrj} warn the reader of the low significance of this feature and that its most plausible explanation is just cosmic variance and/or noise. We can nevertheless use this discussion to argue that features with an amplitude as large as  $30\%$ are still allowed by data. 

However, observation of a peak-like structure would not necessarily imply particle production.  To that end, we computed the bispectrum induced by the particle production. Analogously to the power spectrum, the bispectrum presents a localized enhancement when the three momenta are at the  scale $k \sim a_* H$ which left the horizon during the particle production event.   The existence of the peak again agrees with the results in~\cite{Barnaby:2009dd,Barnaby:2010ke}, which studied non-gaussianity in this model numerically. We provide the analytical scale-dependence of the equilateral bispectrum, and we found that the peak results in a nonlinear parameter $f_{\rm NL,equil} \simeq 3 \cdot 10^6 \, g^{9 / 2}$.  However,  due to its localized nature, observing such a peak would be challenging, but it would be an important step in elucidating the origin of a peak observed in the spectrum.

Furthermore, we have extended this scenario to the case in which particle production occurs in a hidden sector which is only gravitationally coupled to the inflaton. In this case, the full inclusion of metric perturbations is necessary to obtain the phenomenological signatures of the particle production on the inflaton correlators.  The ratio between the sourced and vacuum power spectrum in this scenario also scales as $g^{7 \slash 2}$, with an additional dependence on parameters in the hidden sector such as $|\dot{\psi}_*|$ and $\epsilon_\psi$ which are not constrained by COBE normalization.

To summarize, we have developed a complete framework for evaluating correlation functions in the model~\cite{Chung:1999ve}. We obtained analytic results for the dominant signatures in the power spectrum and bispectrum, and we studied the regime in which such results are under perturbative control. 

\vskip.25cm
\noindent{\bf Acknowledgements:} 

The work of L.P. and M.P. is partially supported from the DOE grant DE-SC0011842  at the University of Minnesota. The work of L.S. is partially supported by the NSF grant PHY-1520292.

\appendix

\section{Evaluation of the one loop integrals } 
\label{app:1loop}

This Appendix is divided into two parts. In the first part we evaluate the dominant one loop diagrams shown in Figure \ref{fig:diagrams},  which account for the modifications of the power spectrum due to particle production. In the second part we compute the modifications of the bispectrum, which are diagrammatically indicated in Figure \ref{fig:diagramsbS}.

\subsection{Power spectrum} 
\label{app:PS}

In this Appendix we evaluate the two expressions (\ref{d1P}) and (\ref{d2P}) which give, respectively, the  one-loop contributions $\delta_1 P_\zeta$ and  $\delta_2 P_\zeta$. Let us start with the first one. Using the $H_{{\rm int},1}^{(0)}$ term in (\ref{Hint-12}), along with the two decompositions (\ref{phi-deco}) and (\ref{chi-deco}), and performing two commutators gives 
\begin{eqnarray}
&& \!\!\!\!\!\!\!\!  \!\!\!\!\!\!\!\!  
\delta_1 \left\langle \zeta \left( \tau ,\, \vec{k} \right)  \zeta \left( \tau ,\, \vec{k}' \right) \right\rangle = \nonumber\\ 
&& \!\!\!\!\!\!\!\!  \!\!\!\!\!\!\!\! 
=  -  \frac{{\cal H}^2}{\varphi^{' 2}}  \, \int^{\tau} d \tau_1  a^4 \left( \tau_1 \right) \,  g^2 \left( \varphi \left( \tau_1 \right)  - \phi_* \right)  \int^{\tau_1} d \tau_2   a^4 \left( \tau_2 \right) \,  g^2 \left( \varphi \left( \tau_2 \right)  - \phi_* \right) 
\int \frac{d^3 p_2 d^3 p_3  d^3 q_2 d^3 q_3 }{\left( 2 \pi \right)^3} \nonumber\\ 
&&   \times 
2  \left[  \delta \phi^{(0)} \left( \tau ,\, \vec{k} \right) ,\,  \delta \phi^{(0)} \left( \tau_1 ,\, - \vec{p}_2 - \vec{p}_3 \right) \right] 
 \left[  \delta \phi^{(0)} \left( \tau ,\,   \vec{k}' \right) ,\, \delta \phi^{(0)} \left( \tau_2 ,\, - \vec{q}_2 - \vec{q}_3 \right) \right] \nonumber\\ 
&&   \times 
 \left\langle  \chi^{(0)} \left( \tau_2 ,\, \vec{q}_2 \right)   \chi^{(0)} \left( \tau_1 ,\, \vec{p}_2 \right) \right\rangle 
\left\langle  \chi^{(0)} \left( \tau_2 ,\, \vec{q}_3 \right)  \chi^{(0)} \left( \tau_1 ,\, \vec{p}_3 \right) \right\rangle  + \left( \vec{k} \, \leftrightarrow \, \vec{k}' \right) \nonumber\\ 
&& \!\!\!\!\!\!\!\! 
 -  \frac{{\cal H}^2}{\varphi^{' 2}}  \, \int^{\tau} d \tau_1  a^4 \left( \tau_1 \right) \,  g^2 \left( \varphi \left( \tau_1 \right)  - \phi_* \right)  \int^{\tau_1} d \tau_2   a^4 \left( \tau_2 \right) \,  g^2 \left( \varphi \left( \tau_2 \right)  - \phi_* \right) 
\int \frac{d^3 p_2 d^3 p_3  d^3 q_2 d^3 q_3 }{\left( 2 \pi \right)^3} \nonumber\\ 
&&   \times 
2  \left[  \delta \phi^{(0)} \left( \tau ,\, \vec{k} \right) ,\,  \delta \phi^{(0)} \left( \tau_1 ,\, - \vec{p}_2 - \vec{p}_3 \right) \right] 
\left\langle  \delta \phi^{(0)} \left( \tau ,\,   \vec{k}' \right)   \delta \phi^{(0)} \left( \tau_2 ,\, - \vec{q}_2 - \vec{q}_3 \right) \right\rangle \nonumber\\ 
&&   \times  \left[  \chi^{(0)} \left( \tau_1 ,\, \vec{p}_2 \right) ,\,  \chi^{(0)} \left( \tau_2 ,\, \vec{q}_2 \right) \right] 
\left\langle  \chi^{(0)} \left( \tau_2 ,\, \vec{q}_3 \right)  \chi^{(0)} \left( \tau_1 ,\, \vec{p}_3 \right) \right\rangle  + \left( \vec{k} \, \leftrightarrow \, \vec{k}' \right)  \,. 
\label{app-start-explicit-1loop}
\end{eqnarray} 

The second term is proportional to the commutator between two $\chi^{(0)}$ fields. We express it as 
\begin{eqnarray}
&& \left[  \chi^{(0)} \left( \tau_1 ,\, \vec{p}_2 \right) ,\,  \chi^{(0)} \left( \tau_2 ,\, \vec{q}_2 \right) \right] = 
\left\langle \left[  \chi^{(0)} \left( \tau_1 ,\, \vec{p}_2 \right) ,\,  \chi^{(0)} \left( \tau_2 ,\, \vec{q}_2 \right) \right] \right\rangle \nonumber\\ 
&& \quad\quad\quad\quad  
\left\langle :  \chi^{(0)} \left( \tau_1 ,\, \vec{p}_2 \right) \,  \chi^{(0)} \left( \tau_2 ,\, \vec{q}_2 \right) - 
\chi^{(0)} \left( \tau_2 ,\, \vec{q}_2 \right) \,   \chi^{(0)} \left( \tau_1 ,\, \vec{p}_2 \right) : \right\rangle = 0 \,, 
\label{eq:simple_chis}
\end{eqnarray}
where the last step follows immediately from (\ref{:chi-chi:}). The vanishing of the commutator amounts to the fact that, after the normal ordering, the quanta of $\chi$ act as a classical source. 

Using again (\ref{phi-deco}) and (\ref{chi-deco}), as well as the the normal ordered product  (\ref{:chi-chi:}), the first term gives 
\begin{eqnarray}
\delta_1 P_\zeta \left( \tau ,\, k \right) &=& - \frac{k^3}{2 \pi^2} \,  \frac{{\cal H}^2}{\varphi^{' 2}}  \, \frac{1}{2}  
 \, \int_{\tau_*}^{\tau} d \tau_1 \, a^2 \left( \tau_1 \right) \,  \frac{g^2 \left( \varphi \left( \tau_1 \right)  - \phi_* \right) }{\omega \left( \tau_1 \right)} \int_{\tau_*}^{\tau_1} d \tau_2 \,  a^2 \left( \tau_2 \right) \,  \frac{g^2 \left( \varphi \left( \tau_2 \right)  - \phi_* \right)  }{\omega \left( \tau_2 \right)} 
\nonumber\\ 
&&  \!\!\!\!\!\!\!\!  \!\!\!\!\!\!\!\!  \!\!\!\!\!\!\!\!  \!\!\!\!\!\!\!\! 
 \times \int \frac{d^3 p_2 d^3 p_3   }{\left( 2 \pi \right)^3} 
 \left( \delta \phi_{k}^{(0)} \left( \tau \right) \, \delta \phi_{k}^{(0)*} \left( \tau_1 \right) - {\rm c.c.} \right)  \delta^{(3)} \left( \vec{k} - \vec{p}_2 - \vec{p}_3 \right) 
\;
   \left( \delta \phi_{k}^{(0)} \left( \tau \right) \, \delta \phi_{k}^{(0)*} \left( \tau_2 \right) - {\rm c.c.} \right) \nonumber\\ 
&& \times 
\, \left[ \vert \beta_{p_2} \vert^2 \Phi \left( \tau_2 \right)  \Phi^* \left( \tau_1 \right) 
+ \alpha_{p_2} \beta_{p_2}^* \Phi^* \left( \tau_2 \right)  \Phi^* \left( \tau_1 \right) + {\rm c.c.} \right] \nonumber\\ 
&& \times
\, \left[ \vert \beta_{p_3} \vert^2 \Phi \left( \tau_2 \right)  \Phi^* \left( \tau_1 \right) 
+ \alpha_{p_3} \beta_{p_3}^* \Phi^* \left( \tau_2 \right)  \Phi^* \left( \tau_1 \right) + {\rm c.c.} \right] \nonumber\\ 
&&  + \left( \vec{k} \, \leftrightarrow \, - \vec{k} \right) \;. 
\end{eqnarray} 

We disregard the contribution of the momentum in the phases $\Phi$, as we have seen in the main text that the quanta are highly non-relativistic. 
We then perform the multiplication, and we drop the oscillating terms (namely, terms in which $\Phi$ is still present), to obtain 
\begin{eqnarray}
\delta_1 P_\zeta \left( \tau ,\, k \right) &\simeq& 
 - \frac{k^3}{2 \pi^2} \,  \frac{{\cal H}^2}{\varphi^{' 2}}  \, \frac{g^2}{2}   \, \int_{\tau_*}^{\tau} d \tau_1  a \left( \tau_1 \right) \, 
\int_{\tau_*}^{\tau_1} d \tau_2   a \left( \tau_2 \right) \,   
\int \frac{d^3 p_2 d^3 p_3   }{\left( 2 \pi \right)^3} \nonumber\\ 
&& \!\!\!\!\!\!\!\!  \!\!\!\!\!\!\!\! \times 
 \left( \delta \phi_{k}^{(0)} \left( \tau \right) \, \delta \phi_{k}^{(0)*} \left( \tau_1 \right) - {\rm c.c.} \right)  \delta^{(3)} \left( \vec{k} - \vec{p}_2 - \vec{p}_3 \right) 
\;    \left( \delta \phi_{k}^{(0)} \left( \tau \right) \, \delta \phi_{k}^{(0)*} \left( \tau_2 \right) - {\rm c.c.} \right) \nonumber\\ 
&& \!\!\!\!\!\!\!\!  \!\!\!\!\!\!\!\! \times 
\left[ \alpha_{p_2} \alpha_{p_3}^* \beta_{p_2}^* \beta_{p_3} +  \alpha_{p_2}^* \alpha_{p_3} \beta_{p_2} \beta_{p_3}^* + 
2 \vert \beta_{p_2} \vert^2  \vert \beta_{p_3} \vert^2  \right]  + \left( \vec{k} \, \leftrightarrow \, - \vec{k} \right) \,. 
\label{eq:ref}
\end{eqnarray} 

As is clear from (\ref{alpha-beta}), and as we will find shortly, the momentum integrals are dominated by the internal (comoving) momenta $p_i \sim \sqrt{g \vert \dot{\varphi}_* \vert} a_* = a_* / \delta t_*$ (where $\delta t_*$ is the time during which the production of $\chi-$quanta takes place,  as discussed in the main text after eq. (\ref{adiabaticity})). On the other hand, as we will see, the correction that we are computing has a peak at values of the (comoving) external momentum $k \sim a_* H$. Quick particle production requires $\delta t_* \, H \ll 1$ (as discussed in the main text), which is a condition that we impose. As a consequence, we can eliminate the external momentum in comparison with the internal one in the arguments of the  Bogolyubov coefficients, and find 
\begin{eqnarray}
&& \int \frac{d^3 p_2 d^3 p_3   }{\left( 2 \pi \right)^3} \;   \delta^{(3)} \left( \vec{k} - \vec{p}_2 - \vec{p}_3 \right) \; \left[ \alpha_{p_2} \alpha_{p_3}^* \beta_{p_2}^* \beta_{p_3} +  \alpha_{p_2}^* \alpha_{p_3} \beta_{p_2} \beta_{p_3}^* + 2 \vert \beta_{p_2} \vert^2  \vert \beta_{p_3} \vert^2  \right]  \nonumber\\ 
&& \quad\quad \simeq \int \frac{d^3 p  }{\left( 2 \pi \right)^3}  \,  
\left[ 2 \vert \alpha_p \vert^2 \, \vert \beta_p \vert^2 + 2 \vert \beta_p \vert^4  \right]  = \frac{2+\sqrt{2}}{8 \pi^3} \, a_*^3 \left( g \vert \dot{\varphi}_* \vert \right)^{3/2} \;. 
\end{eqnarray} 
The $ \left( \vec{k} \, \leftrightarrow \, - \vec{k} \right) $ permutation then just results in a factor of $2$, and we find 
\begin{eqnarray}
\delta_1 P_\zeta \left( \tau ,\, k \right) &\simeq& 
 \frac{2+\sqrt{2}}{16 \pi^5} \, \frac{a_*^3}{k^3} \left( g \vert \dot{\varphi}_* \vert \right)^{3/2} \, \frac{H^2}{\dot{\varphi}^2_*} 
   \,   g^2  \, H^2     \nonumber\\ 
&& \!\!\!\!\!\!\!\!  \!\!\!\!\!\!\!\! 
 \int_{\tau_*}^{\tau} \frac{d \tau_1}{ -\tau_1} \,  \left[ k \tau_1 \cos \left( k \tau_1 \right) - \sin \left( k \tau_1 \right) \right] \; 
  \int_{\tau_*}^{\tau_1} \frac{d \tau_2}{ -\tau_2} \,  \left[ k \tau_2 \cos \left( k \tau_2 \right) - \sin \left( k \tau_2 \right) \right]\,.  
\end{eqnarray} 

The two time integrals can be symmetrized. Moreover, since the integrand is regular at $\tau =0$, and since we are interested in the super horizon $-k \tau \ll 1$ limit, we can simply set $\tau=0$ in the extreme of integration, and write, with a change of variable, 
\begin{equation} 
\delta_1 P_\zeta \left( \tau ,\, k \right) \Bigg\vert_{-k \tau \ll 1} \simeq \frac{2+\sqrt{2}}{16 \pi^5} \, \frac{a_*^3}{k^3} \left( g \vert \dot{\varphi}_* \vert \right)^{3/2} 
   \,   g^2 \,  \frac{ H^4 }{\dot{\varphi}_*^2}   \;\;  \frac{1}{2} \, \left\{ 
 \int_0^{x_*} \frac{d x_1}{ x_1} \,  \left[ x_1 \cos \left( x_1 \right) - \sin \left( x_1 \right) \right] \right\}^2  \;, 
\label{app-end-explicit-1loop}
\end{equation} 
where $x_* \equiv - k \, \tau_* = \frac{k}{a_* H}$. Performing the time integration and dividing by the unperturbed power spectrum gives the first result in (\ref{P-int-res}).

Let us now move to the computation of the second loop diagram. Proceeding as with the first diagram, we obtain 
\begin{eqnarray} 
&& \!\!\!\!\!\!\!\!  \!\!\!\!\!\!\!\! 
\delta_2 \left\langle \zeta \left( \vec{k} \right)  \zeta \left( \vec{k}' \right) \right\rangle = 
- i \; \frac{g^2}{2} \;   \frac{{\cal H}^2}{\varphi^{' 2}} \,  \left( \varphi \left( \tau \right) \right) 
\int \frac{d^3 p_1 \; d^3 p_2 \; d^3 q_1 \; d^3 q_2}{\left( 2 \pi \right)^3} \; 
\delta^{(3)} \left( \vec{p}_1 +   \vec{p}_2 +   \vec{q}_1 +   \vec{q}_2 \right) \int^\tau  d \tau_1 \; a^4 \left( \tau_1 \right) \nonumber\\ 
&&  \quad\quad \times 
\left\langle : \chi^{(0)} \left( \tau_1 ,\, \vec{q}_1 \right) \; \chi^{(0)} \left( \tau_1 ,\, \vec{q}_2 \right) : \right\rangle \; 
\left\langle \left[ \delta \phi^{(0)} \left( \tau ,\, \vec{k} \right) \; \delta \phi^{(0)} \left( \tau ,\, \vec{k}' \right) \;, 
\delta \phi^{(0)} \left( \tau_1 ,\, \vec{p}_1 \right) \; \delta \phi^{(0)} \left( \tau_1 ,\, \vec{p}_2 \right)  \right] \right\rangle \,. \nonumber\\ 
\end{eqnarray} 
We again use (\ref{:chi-chi:}), dropping the fast oscillating terms, and we disregard the external momenta in comparison with the internal ones. 
This leads to 
\begin{eqnarray} 
&& \!\!\!\!\!\!\!\!  \!\!\!\!\!\!\!\! 
\delta_2 \left\langle \zeta \left( \vec{k} \right)  \zeta \left( \vec{k}' \right) \right\rangle = 
- i \; \frac{g^2}{2} \;   \frac{{\cal H}^2}{\varphi^{' 2}} \, \left( \varphi \left( \tau \right) \right) \; 
\left( \int  \frac{d^3 q }{\left( 2 \pi \right)^3} \;   \vert \beta_q \vert^2  \right) \; \int d^3 p   \nonumber\\ 
&& 
\int_{\tau_*}^\tau  d \tau_1 \; 
\frac{ a^2 \left( \tau_1 \right) }{  \omega \left( \tau_1 \right)   } 
\left\langle \left[ \delta \phi^{(0)} \left( \tau ,\, \vec{k} \right) \; \delta \phi^{(0)} \left( \tau ,\, \vec{k}' \right) \;, 
\delta \phi^{(0)} \left( \tau_1 ,\, \vec{p} \right) \; \delta \phi^{(0)} \left( \tau_1 ,\, - \vec{p} \right)  \right] \right\rangle \;, \nonumber\\ 
\end{eqnarray} 
where 
\begin{equation}
\int \frac{d^3 q}{\left( 2 \pi \right)^3} \, \vert \beta_q \vert^2 = 
\frac{a^3 \left( \tau_* \right) \, g^{3/2} \vert \dot{\varphi}_* \vert^{3/2}}{8 \pi^3} \;. 
\end{equation} 

Using the decomposition (\ref{phi-deco}) and the mode functions (\ref{df0}) we obtain, after some algebra, 
\begin{eqnarray} 
&& \!\!\!\!\!\!\!\!  \!\!\!\!\!\!\!\! 
\delta_2 P_\zeta \left( \tau ,\, k \right)  \simeq 
 \; \frac{H^2}{\dot{\varphi}_*^2} \,  \left( \varphi \left( \tau \right) \right) \; 
\frac{a^3 \left( \tau_* \right) \, g^{7/2} \vert \dot{\varphi}_* \vert^{3/2} H^3}{32 \pi^5 k^3} \nonumber\\ 
&& \quad\quad \quad\quad \quad\quad 
\times   \int_{-k \tau}^{-k \tau_*}  \frac{d \left( - k \tau_1 \right)}{-k \tau_1} \; 
\frac{ 
- 2 k \tau_1 \, \cos \left( 2 k \tau_1 \right) + \left( 1 - k^2 \tau_1^2 \right)  \sin \left( 2 k \tau_1 \right) 
 }{ g \vert \varphi \left( \tau_1 \right) - \phi_* \vert   } \;. 
\label{d2Pz-partial}
\end{eqnarray} 

The integral is highly dominated by the earliest times $\tau_1 \simeq \tau_*$ right after the particle production, where, formally, the denominator vanishes. At these times we can approximate 
\begin{equation}
\vert \varphi \left( \tau_1 \right) - \varphi_* \vert \simeq \vert \dot{\varphi}_* \vert \; \left( t - t_* \right) = \frac{ \vert \dot{\varphi}_* \vert }{H} \; \ln \left( \frac{a \left( \tau_1 \right)}{a \left( \tau_* \right)} \right) =  \frac{ \vert \dot{\varphi}_* \vert }{H} \; \ln \left( \frac{- k \, \tau_*}{-k \, \tau_1}  \right)  \;. 
\label{mass-app} 
\end{equation} 
The contribution at $\tau_1 \simeq \tau$ is negligible, and, since we are interested in the super-horizon $-k \tau \ll 1$ limit, we can simply set the lower integration limit to zero. We then denote the integration variable by $x=- k \tau_1$ and, as before, we define $x_* = - k \, \tau_*$. 
We can bring outside of the integral all terms that are regular at $\tau_*$, and write 
\begin{eqnarray} 
&& \!\!\!\!\!\!\!\!  \!\!\!\!\!\!\!\! 
\delta_2 P_\zeta \left( \tau ,\, k \right)   \Bigg\vert_{-k \tau \ll 1} \simeq   \frac{H^2}{\dot{\varphi}_*^2} \, 
\frac{a^3 \left( \tau_* \right) \, g^{5/2} \vert \dot{\varphi}_* \vert^{1/2} H^4}{32 \pi^5 k^3} \;
\frac{ 2 x_* \,  \cos \left( 2 x_* \right) - \left( 1 - x_*^2 \right)  \sin \left( 2 x_* \right) 
 }{ x_*   } \nonumber\\
 &&\qquad\qquad\qquad \times  \int_0^{x_* \left( 1 - \epsilon \right)}  \frac{d x}{  \ln \left( \frac{x_*}{x}  \right)  } \;, 
\label{d2Pz-partial2}
\end{eqnarray} 
where 
\begin{equation}
\epsilon = \left\vert  \frac{ \delta \tau_* }{ \tau_* } \right\vert =  \left\vert  \frac{ \delta t_* }{ a \left( \tau_* \right) \, \tau_* } \right\vert = H \; \delta t_* = 
 \frac{H}{\sqrt{g \, \vert \dot{\varphi} \vert_*}} \;. 
\end{equation} 
Namely, we use the fact that the expressions (\ref{alpha-beta}) are actually valid at $\tau_* + \delta \tau_*$ (after the particle production has ceased), corresponding to $x = x_* \left( 1 - \epsilon \right)$. As we discussed in the main text after eq. (\ref{adiabaticity}), we compute the contribution to $\delta_2 P_\zeta$ only from the times after which the particle production has saturated, and we disregard the contribution from the short time at which the production takes place. The divergence in (\ref{d2Pz-partial}) is not real, and it is due to the fact that we have used the approximate expression $\omega \left( \tau_1 \right) \simeq a\, g\, \vert \varphi \left( \tau_1 \right) - \varphi_* \vert$, which is invalid during this short time interval. Once we assume $\tau_1 > \tau_* + \delta \tau_*$, then the approximations we have performed are valid, and we can obtain a reliable estimate for this correction. 

The remaining time integration evaluates to 
\begin{equation}
 \int_0^{x_* \left( 1 - \epsilon \right)}  \frac{d x}{  \ln \left( \frac{x_*}{x}  \right)  }  = x_* \, 
 {\rm LogIntegral } \left( 1 - \epsilon \right)  \simeq x_* \, \ln \epsilon = x_* \, \ln \left(  \frac{H}{\sqrt{g \, \vert \dot{\varphi} \vert_*}} \right) \;. 
\end{equation}
Inserting this result in (\ref{d2Pz-partial2}), and dividing by the unperturbed power spectrum, gives the second result in (\ref{P-int-res}).

\subsection{Bispectrum } 
\label{app:BS}

In this section we discuss the calculation of the bispectrum, particularly equation \eqref{eq:BS}.  Within the in-in formalism, the left graph of Fig.~\ref{fig:corrections3} corresponds to the expectation value
\begin{align}
&\left< \zeta (\tau, \vec{k}_1)  \zeta(\tau, \vec{k}_2) \zeta(\tau, \vec{k}_3) \right>_{I}
= - \dfrac{H^3}{|\varphi_*|^3} (-i)^3 \int^\tau d\tau_1 \int^{\tau_1} d\tau_2 \int^{\tau_2} d\tau_3  \nonumber \\
&\qquad \left< [[[ \delta \phi^{(0)}(\tau,\vec{k}_1) \, \delta \phi^{(0)}(\tau,\vec{k}_2) \, \delta \phi^{(0)}(\tau,\vec{k}_3), H^{(0)}_{\mathrm{int},1}(\tau_1)],H^{(0)}_{\mathrm{int},1}(\tau_2)],H^{(0)}_{\mathrm{int},1}(\tau_3)]\right>.
\end{align}
After substituting the vertices, we note that the $\chi^{(0)}$ operators give trivial contributions to the commutations, and can be collected in a  $\left< :\chi^6: \right>$ correlator. As in the computation of the power spectrum, this correlator can be split in products of two point functions (\ref{:chi-chi:}) exploiting the gaussianity of the $\chi-$field.  This gives 
\begin{align}
&\!\!\!\!\!\!\!\!\!\!\!\! \left< \zeta (\tau, \vec{k}_1)  \zeta(\tau, \vec{k}_2) \zeta(\tau, \vec{k}_3) \right>_{I} = - \dfrac{i g^3 H^3}{|\dot{\varphi}_*|^3}
\int^\tau_{\tau_*} d\tau_1 \, a(\tau_1) 
\int^{\tau_1}_{\tau_*} d\tau_2 \, a(\tau_2) 
\int^{\tau_2}_{\tau_*} d\tau_3 \, a(\tau_3) \nonumber \\
& \int \dfrac{d^3 p_1 \, d^3 p_2\, d^3 \ell_1 }{(2\pi)^{9 \slash 2}} 
\left< [[[ \delta \phi^{(0)}(\tau,\vec{k}_1) \, \delta \phi^{(0)}(\tau,\vec{k}_2) \, \delta \phi^{(0)}(\tau,\vec{k}_3), \delta \phi^{(0)}(\tau_1,-\vec{p}_1 - \vec{p}_2)], \delta \phi^{(0)}(\tau_2,\vec{\ell}_1 + \vec{p}_1) ], \right. \nonumber \\
& \qquad \qquad \left. \delta \phi^{(0)}(\tau_3,-\vec{\ell}_1 + \vec{p}_2) ] \right> 
\nonumber \\
&
\cdot \beta_{\ell_1} \beta_{p_1} \beta_{p_2}  
\left(\alpha_{\ell_1} \beta_{p_1} \alpha^*_{p_2}+\alpha_{p_2} (\beta_{\ell_1} \alpha^*_{p_1}+\alpha^*_{\ell_1} \beta_{p_1})+\alpha_{\ell_1} \alpha^*_{p_1} \beta_{p_2}+\alpha_{p_1} (\beta_{\ell_1} \alpha^*_{p_2}+\alpha^*_{\ell_1} \beta_{p_2})+2 \beta_{\ell_1} \beta_{p_1} \beta_{p_2}\right),
\end{align} 
where we have noted that in equation \eqref{alpha-beta} we have chosen $\beta$ to be real.  We have dropped all terms which retain a quickly oscillating phase.  Next we expand the nested commutator of the $\delta \phi^{(0)}$ fields to find, for the exact equilateral configuration,
\begin{align}
&\!\!\!\!\!\!\!\!\!\!\!\!\left< \zeta (\tau, \vec{k}_1)  \zeta(\tau, \vec{k}_2) \zeta(\tau, \vec{k}_3) \right>^\prime_{I} \Big\vert_{k_1=k_2=k_3 \equiv k} = \dfrac{ 6 i H^3 g^3}{|\dot{\varphi}|^3}
\int^\tau_{\tau_*} d\tau_1 \, a(\tau_1) 
\int^{\tau_1}_{\tau_*} d\tau_2 \, a(\tau_2) 
\int^{\tau_2}_{\tau_*} d\tau_3 \, a(\tau_3) \nonumber \\
&
\left[\varphi^{(0)*}_{k}(\tau ) \varphi^{(0)} _{k}(\tau_1)-c.c.\right] 
\left[\varphi^{(0)*}_{k}(\tau ) \varphi^{(0)} _{k}(\tau_2)-c.c.\right] 
\left[ \varphi^{(0)*}_{k}(\tau ) \varphi^{(0)} _{k}(\tau_3)-c.c.\right] \nonumber \\
& \cdot \int \dfrac{d^3 p_2}{(2\pi)^{9 \slash 2}} 
 \beta_{p_2}^3   
\left(6 |\alpha_{p_2}|^2 \beta_{p_2} +2 \beta_{p_2}^3 \right),
\end{align}
where (as discussed  after equation \eqref{eq:ref})  in the argument of the Bogolyubov coefficients we have used the fact that  the result is dominated by values of the internal momentum much greater than $k$.  The momentum integral can be immediately done using (\ref{alpha-beta}). We then insert the mode functions (\ref{df0}), and evaluate the time integrals in the  $\tau \rightarrow 0$ limit, to obtain 
\begin{align}
\left< \zeta (\tau, \vec{k}_1)  \zeta(\tau, \vec{k}_2) \zeta(\tau, \vec{k}_3) \right>^\prime_{I}
&= \dfrac{H^3}{|\dot{\varphi}_*|^3} \cdot \dfrac{g^3}{k^6  } \cdot  \dfrac{1}{(2\pi)^{9 \slash 2}} \dfrac{(27 \sqrt{2} + 16 \sqrt{3}) (g |\dot{\varphi}_*|)^{3 \slash 2}}{18} \dfrac{\left[ -\sin(x_*) + \mathrm{SI}(x_*) \right]^3}{x_*^3},
\label{eq:BS_1}
\end{align}
where $x_* = - k \tau_*$, which is eq. \eqref{eq:BS} of the main text. 

We also evaluated the contribution to the bispectrum from the right diagram of  Fig.~\ref{fig:corrections3} (we denote this contribution by $\left\langle \zeta^3 \right\rangle_{II}$) and we found that it is suppressed with respect to (\ref{eq:BS_1}). Specifically, we found that 
\begin{align}
\left\vert \dfrac{\left< \zeta^3 \right>_{II}}{\left< \zeta^3 \right>_{I}} \right\vert \sim \dfrac{H^2}{g \, \vert \dot{\varphi_*} \vert} \ln \left( \dfrac{\sqrt{ 2\,\epsilon_*} g M_p}{H} \right) \ll 1 \,. 
\end{align}
This ratio can be readily obtained from the Feynman rules outlined  in Appendix \ref{app:Feynman}.  Therefore, the diagram with all cubic vertices dominates the bispectrum, in analogy to what we have found  for the power spectrum.

\section{Feynman rules and perturbativity discussion } 
\label{app:Feynman}

As we show in Appendix~\ref{app:zeta} and \ref{app:dg}  below,  knowledge of the perturbations of the inflaton field  $\delta\phi$ is sufficient to determine with excellent accuracy the spectra of the scalar perturbations in this model. In order to know when to trust our results we must therefore establish the regime of validity of the perturbation series that determines $\delta\phi$. This task is not trivial, as the model has an obvious ``small'' expansion parameter $g$, but it is also characterized by a ``large'' combination $\sqrt{g\,|\dot\varphi_*|}/H $. In this Appendix we will outline the rules that allow us to determine how the various contributions to $\delta\phi$, expressed in terms of Feynman diagrams, scale with $g$ and with $\sqrt{g\,|\dot\varphi_*|}/H$, and we deduce the conditions for the validity of the perturbative expansion. 

The vertices of the theory are given by eqs.~(\ref{Hint-12}). The second of those equations determines a cubic vertex $\delta\phi-\chi-\chi$ with coefficient $g^2\,\left[\varphi(\tau)-\phi_*\right]\simeq g\,(g\,|\dot\phi_*|/H)$  (as the signature of the $\chi-$quanta is imprinted in the first ${\rm O } \left( 1 \right)$ Hubble times after they are generated, before they redshift away), whereas the third equation gives a $\delta\phi-\delta\phi-\chi-\chi$ quartic vertex with coefficient $g^2$. Moreover, loops of the $\chi$ field come with a measure $d^3 q/(2\pi)^3$ and are cutoff by the ultraviolet behavior of the Bogolyubov coefficients at $q\simeq \sqrt{g\,|\dot\varphi_*|}$, therefore giving a factor $(g\,|\dot\varphi_*|)^{3/2}/(2\pi)^3$. Finally, propagators of the $\chi$ fields get, as one can see from see eq.~(\ref{:chi-chi:}), a factor $(\sqrt{\omega})^{-2}\simeq (g\,|\dot\varphi_*|/H)^{-1}$.

Therefore, to sum up, the scaling of each Feynman diagram will be given by the following rules:

\begin{enumerate}

\item each cubic vertex gives a factor $g\,(g|\dot\varphi_*|/H^2)$;

\item each quartic vertex gives a factor $g^2$;

\item each loop of $\chi$ gives a factor $(g|\dot\varphi_*|/H^2)^{3/2}/(2\pi)^3$;

\item each $\chi$ propagator gives a factor $(g|\dot\varphi_*|/H^2)^{-1}$;

\item since the rules above give all dimensionless factors, we have to add factors of $H$ to make the correlator dimensionally consistent. 

\end{enumerate}

Let us note here that we will not be concerned here with possible infrared divergences that might give rise to additional $\log\left(\sqrt{g\,|\dot\varphi_*|}/H\right)$ factors.  Moreover, these rules do not give the momentum dependence of the correlators. Invariance under the de Sitter symmetries implies that the two point functions, once one factors out a measure $\delta(\vec{k}+\vec{k}')/k^3$, depend on momentum and $\tau_*$ only through the dimensionless combination $k\,\tau_*$.  

In order to check the validity of our rules let us observe that, when we apply them to the diagrams of Figure~\ref{fig:diagrams}, we obtain respectively $H^2\times\left[g\,(g|\dot\varphi_*|/H^2)\right]^2\times[(g|\dot\varphi_*|/H^2)^{3/2}/(2\pi)^3]\times\left[(g|\dot\varphi_*|/H^2)^{-1}\right]^2\sim \frac{g^{7/2}}{(2\pi)^3}\,|\dot\varphi_*|^{3/2}/H$ and $H^2\times g^2\times[(g|\dot\varphi_*|/H^2)^{3/2}/(2\pi)^3]\times (g|\dot\varphi_*|/H^2)^{-1}\sim \frac{g^{5/2}}{(2\pi)^3}\,|\dot\varphi_*|^{1/2}\,H$. Those results, once divided by $H^2$ to normalize to the unperturbed two point function of the $\delta\phi$, give scalings that agree  with eqs.~(\ref{P-int-res}). We have also checked the validity of the above scalings on five more one- and two-loop corrections to the one- and two-point function of $\delta\phi$, as well as for two diagrams discussed in Subsection~\ref{subsec:BS} and Appendix \ref{app:BS} that contribute to the three point function of $\delta\phi$.

We study the convergence  of the Feynman series for the two point function of $\delta\phi$,  by requiring that two-loop corrections are subdominant with respect to the leading one-loop contribution $\sim \frac{g^{7/2}}{(2\pi)^3}\,|\dot\varphi_*|^{3/2}/H$, given by the first  diagram in Figure~\ref{fig:diagrams}. We evaluated all the two loop diagrams, and found that the dominant one is a chain diagram obtained by attaching with a $\delta \phi$ propagator the two diagrams in the left and right panel of Figure~\ref{fig:diagrams}. (A direct computation shows that instead a chain diagram obtained from attaching two copies of the left panel of  Figure~\ref{fig:diagrams} vanishes.) The exact computation of this diagram is well estimated by the Feynman rules listed above. We obtain  $\sim \left(\frac{g^{7/2}}{(2\pi)^3}\,|\dot\varphi_*|^{3/2}\right)\times \left(\frac{g^{5/2}}{(2\pi)^3}\,|\dot\varphi_*|^{1/2}\right)$, so that convergence of our series will require $g^{5/2}\,|\dot\varphi_*|^{1/2}/H\ll (2\pi)^3$. Setting  $\sqrt{|\dot\varphi_*|}/H\simeq 58$, as required to obtain the correct normalization of the power spectrum  $P_\zeta \simeq P_\zeta^{(0)} \simeq 2.2 \cdot 10^{-9} $  \cite{Ade:2015lrj}, this turns into the condition
\begin{equation}
g^2\lesssim \left(\frac{(2\pi)^3\,H}{\sqrt{|\dot\varphi_*|}}\right)^{4/5}\simeq 3 \,. 
\label{perturbativity} 
\end{equation}

This condition, even if it eventually leads to the requirement $g^2\lesssim {\cal O}(1)$ that one would have naively guessed, does actually emerge from a nontrivial competition of the two large factors $(2\pi)^3$ and $\sqrt{|\dot\varphi_*|}/H$. It also implies that the leading order correction $\sim \frac{g^{7/2}}{(2\pi)^3}\,|\dot\varphi_*|^{3/2}/H$ to the two-point function of the inflaton can be parametrically larger that the ``leading'' vacuum contribution $\sim (H/2\pi)^2$ while being under perturbative control.~\footnote{At two loops and beyond two-loops, we verified that all chain diagrams that have more than two cubic vertices vanish. We interpret these diagrams as two $\delta \phi$ quanta generated by the rescattering of $\chi$ quanta against the inflaton zero mode; after they are generated, these quanta have additional $4-$point interactions from the $\delta \chi^2 \delta \phi^2$ vertex, but do not return into the inflaton condensate. The absence of these processes, and the fact that the one-loop arising from the quartic vertex is smaller than that from the cubic vertices, is the reason why we can have a dominant one-loop diagram. 
} 

Before concluding this Appendix let us note that there is also a {\em lower} bound on the coupling $g$ that originates from requiring that the mass of $\chi$ evolves quickly enough during the event of particle production, $\delta t_*\ll H^{-1}$. This requirement reads $g\,\dot\phi_*\gg H^2$, implying
\begin{align}
g^2\gg 10^{-7}\,.
\end{align}

\section{Justification of $\zeta = - \frac{H}{\dot{\varphi} } \delta \phi$ } 
\label{app:zeta}

In this Appendix we justify the relation $\zeta = - \frac{H}{\dot{\varphi}} \; \delta \phi$ used in the main text. We perform our computation in spatially flat gauge, characterized by the line element 
\begin{equation}
d s^2 = a^2 \left( \tau \right) \left[ - \left( 1 + 2 \Phi \right) d \tau^2 + \partial_i B \, d \tau \, d x^i + dx^i d x^i \right] \;, 
\label{line} 
\end{equation} 
where we have ignored vector and tensor metric perturbations. In this Appendix we actually disregard the metric perturbations $\Phi$ and $B$. In the next Appendix we show that this is an accurate approximation. 

Expressed in this gauge, the gauge invariant perturbation $\zeta$ reads 
\begin{equation}
\zeta \left( \tau,\, \vec{x} \right) = - \frac{{\cal H} \left( \tau \right)}{\rho_{\rm bck}' \left( \tau \right)} \, \delta \rho \left( \tau ,\, \vec{x} \right) \;. 
\label{zeta-exact}
\end{equation} 
with  
\begin{equation}
\rho_{\rm bck } \equiv \langle \rho \rangle \;\;,\;\; 
\delta \rho \equiv \rho - \rho_{\rm bck } \;. 
\end{equation} 
where $ \langle X \rangle $ is the theoretical expectation value of the operator $X$. Ignoring metric perturbations, the energy density of the model is given by 
\begin{equation}
\rho = - T^0_0 =  \frac{1}{a^2} \left[  \frac{1}{2 } \phi'  \phi' +  \frac{1}{2 } \partial_i \phi \partial_i  \phi +  \frac{1}{2 } \chi'  \chi' +  \frac{1}{2 } \partial_i \chi \partial_i  \chi + a^2 \, V \left( \phi \right) + a^2 \, \frac{g^2}{2} \left( \phi - \phi_* \right)^2 \chi^2 \right] \;. 
\label{rho-dg=0}
\end{equation} 

We divide the remainder of this Appendix into two parts. In the first one we compute the unperturbed power spectrum of $\zeta$. In the second part we study the leading  corrections to the spectrum in the in-in formalism. We show that in both cases, the assumption $\zeta = - \frac{\cal H}{\varphi'} \; \delta \phi$ made in the main text provides an accurate approximation of the exact relation (\ref{zeta-exact}).

\subsection{Unperturbed $\zeta$} 
\label{app:zeta0}

Here we work to zeroth order in the interaction hamiltonian (\ref{Hint}). We first evaluate 
\begin{eqnarray}
 \rho_{\rm bck}^{(0)'} &=&  \frac{1}{a^2} \left\langle 
 - 3 \frac{a'}{a} \phi^{{(0)}' 2}  - \frac{a'}{a} \partial_i \phi^{(0)} \partial_i \phi^{(0)}   - 3 \frac{a'}{a} \chi^{{(0)}' 2}  - \frac{a'}{a} \partial_i \chi^{(0)} \partial_i \chi^{(0)}  \right\rangle  \nonumber\\ 
 &  \simeq &  \frac{1}{a^2} \left[  - 3 \frac{a'}{a} \varphi^{' 2} + 
 \left\langle 
  - 3 \frac{a'}{a} \chi^{{(0)}' 2}  - \frac{a'}{a} \partial_i \chi^{(0)} \partial_i \chi^{(0)}  \right\rangle  \right] \;, 
\label{rhodot-0}
\end{eqnarray}
which is the background expectation value of the time derivative of (\ref{rho-dg=0}).  In the second line we disregard the contribution from the inflaton perturbations to the expectation value. To evaluate the $\chi-$contribution we proceed identically to the way we did in the main text to derive equation (\ref{rho-chi-bck}), and obtain 
\begin{equation} 
 \rho_{\rm bck}^{(0)'} \simeq  - 3 \, {\cal H} \, \left[ \dot{\varphi}^2 + \rho_{\chi,{\rm bck}}  \right] \;, 
\label{rhodot-0-res}
 \end{equation} 
in agreement with the fact that the energy density of $\chi-$quanta redshifts as that of matter in the adiabatic regime. 
Due to the redshift, the  $\chi-$dependent contribution becomes negligible soon after the particle production.

Next, we move to the computation of 
\begin{equation}
\left\langle \delta \rho^{(0)} \left( \vec{k} \right) \, \delta \rho^{(0)} \left( \vec{k}' \right) \right\rangle = 
\left\langle \rho^{(0)} \left( \vec{k} \right)  \rho^{(0)} \left( \vec{k}' \right) \right\rangle = 
\int \frac{d^3 x d^3 y}{\left( 2 \pi \right)^3} \, {\rm e}^{-i \vec{k} \cdot \vec{x}-i \vec{k}' \cdot \vec{y}} \, 
\left\langle \rho^{(0)} \left( t ,\, \vec{x} \right)  \rho^{(0)} \left( t ,\, \vec{y} \right) \right\rangle \;, 
\end{equation}
where we are implicitly assuming $k,\, k' \neq 0$, so that $ \delta \rho^{(0)} \left( \vec{k} \right) =  \rho^{(0)} \left( \vec{k} \right) $. As we are interested in modes in the super-horizon regime, we disregard the contribution to the energy from the gradients of the fields. We also disregard the contribution to the energy from the kinetic energy of the inflaton, which is slow roll suppressed. Namely, we use 
\begin{equation}
\rho^{(0)} \simeq    \frac{1}{2 a^2} \chi^{(0)'}  \chi^{(0)'}  +  V \left( \varphi + \delta \phi^{(0)} \right) +  \frac{g^2}{2} \left( \varphi - \phi_* \right)^2 \chi^{(0)2}  \;. 
\end{equation}

We have two separate correlators, one for the inflaton and one for $\chi$. For the first one we find 
\begin{eqnarray}
 & & \left\langle \rho^{(0)} \left( \vec{k} \right)  \rho^{(0)} \left( \vec{k}' \right) \right\rangle_\phi =
\int \frac{d^3 x d^3 y}{\left( 2 \pi \right)^3} \, {\rm e}^{-i \vec{k} \cdot \vec{x}-i \vec{k}' \cdot \vec{y}} \, 
\left\langle V \left( \varphi + \delta \phi^{(0)} \left( \vec{x} \right) \right)  V \left( \varphi + \delta \phi^{(0)} \left( \vec{y} \right) \right) \right\rangle \nonumber\\ 
 & & \quad\quad  \quad\quad 
\simeq V^{' 2} \, \int \frac{d^3 x d^3 y}{\left( 2 \pi \right)^3} \, {\rm e}^{-i \vec{k} \cdot \vec{x}-i \vec{k}' \cdot \vec{y}} \, 
\left\langle \delta \phi^{(0)} \left( \vec{x} \right)  \delta \phi^{(0)} \left( \vec{y} \right) \right\rangle 
=  V^{' 2} \, \left\langle  \delta \phi^{(0)} \left( \vec{k} \right)   \delta \phi^{(0)} \left( \vec{k}' \right)  \right\rangle \;, \nonumber\\ 
\end{eqnarray}
where $V' \equiv \frac{\partial V}{\partial \phi} \Big\vert_{\phi = \varphi}$, and where we have disregarded contributions that are higher order in $\delta \phi$. Proceeding as in the main text, the last expression evaluates to 
\begin{eqnarray} 
\left\langle \rho^{(0)} \left( \vec{k} \right)  \rho^{(0)} \left( \vec{k}' \right) \right\rangle_\phi' &\simeq& 
\frac{2 \pi^2}{k^3} \, V^{' 2} \, P_\phi^{(0)} \left( k \right) = 
\frac{2 \pi^2}{k^3} \, V^{' 2} \, \left( \frac{H}{2 \pi} \right)^2 \;. 
\label{drhophi0}
\end{eqnarray}

For the second correlator, using the decomposition (\ref{chi-deco}), we obtain 
\begin{eqnarray}
&& 
 \left\langle \rho^{(0)} \left( \vec{k} \right)  \rho^{(0)} \left( \vec{k}' \right) \right\rangle_\chi =  
 \left\{ \frac{ \partial_{\tau_1}  \partial_{\tau_2}  \partial_{\tau_3}  \partial_{\tau_4}  }{4 a^4} + \frac{g^2}{2 a^2}  \left( \varphi - \phi_* \right)^2 \,  \partial_{\tau_1}   \partial_{\tau_2} + \frac{g^4}{4}  \left( \varphi - \phi_* \right)^4 \right\} \nonumber\\ 
&& \quad\quad \times 
 \int \frac{d^3 p \,d^3 q}{\left( 2 \pi \right)^3}  \left\langle \chi^{(0)} \left( \tau_1 ,\, \vec{p} \right)  \chi^{(0)} \left( \tau_2 ,\, \vec{k} - \vec{p} \right) 
 \chi^{(0)} \left( \tau_3 ,\, \vec{q} \right)  \chi^{(0)} \left( \tau_4 ,\, \vec{k}' - \vec{q} \right) \right\rangle \Bigg\vert_{\tau_i=\tau} \;. 
\nonumber\\ 
\end{eqnarray}

We use gaussianity of $\chi$ to separate the $4-$point function into the sum of three terms, involving the product of two $2-$point functions each. We regulate the two point functions according to (\ref{:chi-chi:}), and obtain 
\begin{eqnarray}
&&
 \left\langle : \rho^{(0)} \left( \vec{k} \right)  \rho^{(0)} \left( \vec{k}' \right) : \right\rangle_\chi =   \int \frac{d^3 p d^3 q}{\left( 2 \pi \right)^3} 
 \frac{\theta \left( \tau - \tau_* \right)  \, \delta^{(3)} \left( \vec{p} + \vec{k}' - \vec{q} \right) 
 \delta^{(3)} \left( \vec{k} - \vec{p} + \vec{q} \right)  }{4 
a^8 \left( \tau \right)  \omega^2 \left( \tau \right) } \nonumber\\  
&& 
 \frac{1}{a^4 \left( \tau \right)} \,  \left\{ \frac{ \partial_{\tau_1}  \partial_{\tau_2}   \partial_{\tau_3}   \partial_{\tau_4}   }{2 } + \omega^2 \left( \tau \right) \,   \partial_{\tau_1}   \partial_{\tau_2}    + \frac{ \omega^4 \left( \tau \right) }{2}  \right\} \nonumber\\ 
&&  \left[ \vert \beta_p \vert^2 \Phi \left( \tau_1 \right)  \Phi^* \left( \tau_4 \right) + \vert \beta_p \vert^2 \Phi^* \left( \tau_1 \right)  \Phi \left( \tau_4 \right)  + \alpha_p \beta_p^* \Phi^* \left( \tau_1 \right)  \Phi^* \left( \tau_4 \right)   + \alpha_p^* \beta_p \Phi \left( \tau_1 \right)  \Phi \left( \tau_4 \right)  \right] \nonumber\\ 
&&  \left[ \vert \beta_q \vert^2 \Phi \left( \tau_2 \right)  \Phi^* \left( \tau_3 \right) + \vert \beta_q \vert^2 \Phi^* \left( \tau_2 \right)  \Phi \left( \tau_3 \right)  + \alpha_q \beta_q^* \Phi^* \left( \tau_2 \right)  \Phi^* \left( \tau_3 \right)   + \alpha_q^* \beta_q \Phi \left( \tau_2 \right)  \Phi \left( \tau_3 \right)  \right]  \Bigg\vert_{\tau_i=\tau} \;. 
\nonumber\\ 
\end{eqnarray} 
We perform the time differentiation accounting for the conditions (\ref{adiabaticity}); in practice, we disregard the adiabatic variation of $\omega$. After performing the differentiation, we set $\tau' = \tau$, and we disregard the terms that are proportional to the fast oscillating phase $\Phi \left( \tau \right)$. As discussed after eq. (\ref{eq:ref}), we can then disregard the external momentum in comparison to the internal one in the Bogolyubov coefficients. This gives 
\begin{eqnarray}
 \left\langle : \rho^{(0)} \left( \vec{k} \right)  \rho^{(0)} \left( \vec{k}' \right) : \right\rangle_\chi' & \simeq  &  
   \frac{\omega^2 \left( \tau \right) }{ a^8 \left( \tau \right)  }   \int \frac{d^3 p }{\left( 2 \pi \right)^3} \; 
 \left[  \vert \alpha_p \vert^2 \, \vert  \beta_p \vert^2 +  \vert \beta_p \vert^4   \right]  \nonumber\\ 
&=&  \theta \left( \tau - \tau_* \right) 
\frac{1}{a^3 \left( \tau \right)} \, \frac{2+ 2 \sqrt{2}}{16 \pi^3} \; 
\frac{\omega^2 \left( \tau \right)  }{ a^2 \left( \tau \right)  }  
 \;  \frac{ a^3 \left( \tau_* \right) }{ a^3 \left( \tau \right) }   g^3 \, \vert \dot{\varphi}_* \vert^{3/2}. 
\label{drhochi0}
\end{eqnarray} 

The addition of the two contributions (\ref{drhophi0}) and (\ref{drhochi0}) gives 
\begin{eqnarray}
\frac{k^3}{2 \pi^2} \left\langle \rho^{(0)} \left( \vec{k} \right)  \rho^{(0)} \left( \vec{k}' \right) \right\rangle' &=& 
\ V^{' 2} \, P_\phi^{(0)} \left( k \right) 
+  \frac{k^3}{a^3 \left( \tau \right)} \, \frac{2+ 2 \sqrt{2}}{4 \pi^2} \; 
\frac{\omega \left( \tau \right)  }{ a \left( \tau \right)  }  \,   \rho_{\chi,{\rm bck}}  \left( \tau \right) \,, 
\label{drho0}
\end{eqnarray}
where the expression (\ref{rho-chi-bck}) has been used in the second term. We note that the $\chi-$dependent contribution is blue, 
and completely negligible with respect to the $\phi-$dependent contribution at super-horizon scales. 

The  $2$-point function of $\zeta^{(0)}$ is obtained from the ratio of (\ref{drho0}) and (\ref{rhodot-0-res}). As we commented after each of the two results, the $\chi-$contribution to both the numerator and denominator is negligible soon after the particle production and at super horizon scale. We then recover the standard result 
\begin{eqnarray}
\left\langle \zeta^{(0)} \left( \vec{k} \right)  \zeta^{(0)} \left( \vec{k'} \right) \right\rangle &=& \frac{{\cal H}^2}{\rho_{\rm bck}^{'2}} \, 
\left\langle \rho^{(0)} \left( \vec{k} \right)  \rho^{(0)} \left( \vec{k'} \right) \right\rangle \simeq \frac{{\cal H}^2 \, V^{'2}}{9 {\cal H}^2 \, \dot{\varphi}^4 } \, \left\langle \delta \phi^{(0)} \left( \vec{k} \right)  \delta \phi^{(0)} \left( \vec{k'} \right) \right\rangle \nonumber\\ 
&\simeq& \frac{ H^2}{\dot{\varphi}^2} \,  \left\langle \delta \phi^{(0)} \left( \vec{k} \right)  \delta \phi^{(0)} \left( \vec{k'} \right) \right\rangle \,, 
\label{zeta0-phi0}
\end{eqnarray}
where the background slow-roll relation $V' \simeq - 3 \, H \, \dot{\varphi}$ has been used in the final expression. This confirms the accuracy 
of the approximated expression  (\ref{zeta-main}) for $\zeta$ used in the main text, at least at the unperturbed level. In the next subsection we show that this continues to be the case also at the perturbative level.

\subsection{Perturbed $\zeta$} 
\label{app:dzeta}

In the previous subsection we have seen that, at the unperturbed level, $\rho_{\rm back}'$ and $\delta \rho$ are strongly dominated by the
inflation kinetic energy and by the fluctuations of the inflaton potential energy, respectively, and that the contributions from the produced quanta of $\chi$ to these expressions become negligible soon after the particle production and at super horizon scales. Since we are performing computations under the assumption that the interactions in (\ref{Hint}) are in the perturbative regime (see Appendix \ref{app:Feynman}), the corrections to the unperturbed expressions cannot reverse this strong hierarchy between the various unperturbed contributions. Therefore, the leading effects due to particle production are the leading corrections from  (\ref{Hint}) to the inflaton kinetic energy, and to the fluctuations of the inflaton potential energy. We compute these two quantities in this subsection. 

\begin{figure}[ht!]
\centerline{
\includegraphics[width=0.4\textwidth,angle=0]{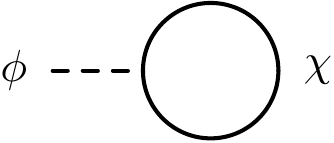}
}
\caption{
Diagrammatic expressions for the leading backreaction from particle production on the motion of the inflaton.}
\label{fig:tadpole}
\end{figure}

We start from the dominant correction to the expectation value of the inflaton field 
\begin{equation}
\delta \left\langle  \phi \left( \tau ,\, \vec{x} \right) \right\rangle = - i \int^\tau d \tau_1 \, \left\langle \left[ \delta \phi^{(0)} \left( \tau ,\, \vec{x} \right) ,\, a^4 \left( \tau_1 \right) g^2 \left( \varphi \left( \tau_1 \right) - \phi_* \right) \int d^3 y \, \delta \phi^{(0)} \left( \tau_1 ,\, \vec{y} \right) 
 \,  \chi^{(0)2} \left( \tau_1 ,\, \vec{y} \right) \right] \right\rangle  \,,  
 \end{equation}
that corresponds to the tadpole diagram shown in Figure \ref{fig:tadpole}. From this expression, we will immediately be able to evaluate the correction to eq. (\ref{rhodot-0-res}).

Using the decompositions (\ref{phi-deco}) and (\ref{chi-deco}), we rewrite 
\begin{eqnarray} 
\delta \left\langle  \phi \left( \tau ,\, \vec{x} \right) \right\rangle &=& - i \, g^2 \, \int^\tau d \tau_1 a^4 \left( \tau_1 \right) \left( \varphi \left( \tau_1 \right) - \phi_* \right) \int \frac{d^3 k \, d^3 p \, d^3 q}{\left( 2 \pi \right)^3} {\rm e}^{i \vec{k} \cdot \vec{x} } \nonumber\\ 
&& \quad \times \left[ \delta \phi^{(0)} \left( \tau ,\, \vec{k} \right) ,\,  \delta \phi^{(0)} \left( \tau_1 ,\, - \vec{p} - \vec{q} \right) \right] \; \left\langle  \chi^{(0)} \left( \tau_1 ,\, \vec{p} \right)   \chi^{(0)} \left( \tau_1 ,\, \vec{q} \right) \right\rangle \,. 
\end{eqnarray} 

Using the expressions (\ref{df0}) and (\ref{:chi-chi:}), we write 
\begin{eqnarray} 
\delta \left\langle  \phi \left( \tau ,\, \vec{x} \right) \right\rangle &=& 2  \, g \, {\rm sign} \left( \dot{\varphi} \right) \, \int_{\tau_*}^\tau d \tau_1 a \left( \tau_1 \right) \,  \int \frac{d^3 k \, d^3 p \, d^3 q}{\left( 2 \pi \right)^3} {\rm e}^{i \vec{k} \cdot \vec{x} } \;  {\rm Im } \left[ \delta \phi_k^{(0)} \left( \tau \right)  \delta \phi_k^{(0)*} \left( \tau_1 \right) \right]  \nonumber\\
&&  \delta^{(3)} \left( \vec{k} + \vec{p} + \vec{q} \right)  \;   {\rm Re } \left[ \vert \beta_p \vert^2 + \alpha_p \beta_p^* \, \Phi^{* 2} \left( \tau_1 \right) \right] \delta^{(3)} \left( \vec{p} + \vec{q} \right)  \,, 
\end{eqnarray} 
and, disregarding the fast oscillating phase, 
\begin{eqnarray} 
\delta \left\langle  \phi \left( \tau ,\, \vec{x} \right) \right\rangle &=& 2  \, g \, {\rm sign} \left( \dot{\varphi} \right)  \, \int_{\tau_*}^\tau \frac{ d \tau_1 }{- H \, \tau_1 } \, \frac{H^2 \left( \tau_1^3-\tau^3 \right)}{6}   \; \int  \frac{ d^3 p }{\left( 2 \pi \right)^3}   \vert \beta_p \vert^2  \nonumber\\ 
&\simeq&  -  \theta \left( \tau - \tau_* \right) \, {\rm sign} \left( \dot{\varphi} \right)  \, 
 \left[ 1 -  \frac{  a^3 \left( \tau_* \right) }{ a^3 \left( \tau \right)} + 3  \frac{  a^3 \left( \tau_* \right) }{ a^3 \left( \tau \right)}  \ln \left( \frac{a \left( \tau_* \right)}{a \left( \tau \right)} \right) \right] \, \frac{  g^{5/2} \vert \dot{\varphi}_* \vert^{3/2}}{72 \pi^3 H^2} \;, \nonumber\\ 
\end{eqnarray} 
Moving to physical time, we obtain  the time derivative  of the inflaton 
\begin{equation} 
\dot{\phi}_{\rm bck} \left( t \right)  \simeq \left\{ \begin{array}{l} 
 \dot{\varphi} \left( t \right) \quad\quad\quad\quad \quad\quad\quad\quad \quad\quad\quad\quad \quad\quad
 ,\;\; \Delta t \equiv t - t_* < 0  \;, \\ 
\dot{\varphi} \left( t \right)  -  \frac{  g^{5/2} \vert \dot{\varphi}_* \vert^{3/2}}{8 \pi^3 }  \, {\rm sign} \left( \dot{\varphi} \right)    \, \Delta t \,   {\rm e}^{-3 H \Delta t}  \;\;,\;\; \Delta t  > 0 \;,
 \end{array} \right. 
\end{equation} 
which is in agreement with the results of~\cite{Chung:1999ve,Romano:2008rr}, which obtained this result by solving the equation of motion for the inflaton zero mode, with the backreaction of the $\chi-$quanta accounted for in the Hartree approximation.  We see that the effect of the backreaction of the produced $\chi$ quanta also redshifts away as $a^{-3}$, and becomes negligible soon after the particle production. This means that the in-in correction to the background energy density (as well as to its derivative and to the Hubble rate) can be disregarded after a quick transient effect immediately following the particle production. 

Let us now turn our attention to the in-in correction to  $\delta \rho$. As we commented at the beginning of this subsection, we are only interested in the correction to fluctuations of the inflaton potential energy. To leading order in the inflaton perturbations, this is given by $\delta^{\rm in-in} V \left( \tau ,\, \vec{k} \right) = \frac{\partial V}{\partial \phi} \Big\vert_{\phi = \varphi \left( \tau \right)} \, \delta^{\rm in-in}  \phi \left( \tau ,\, \vec{k} \right) $, where $\delta^{\rm in-in} X$ denotes the  correction to a quantity $X$ in  the in-in computation. 

Based on this, the relation (\ref{zeta0-phi0}) is extended to 
\begin{eqnarray}
\zeta \left( k \right) &=& - \frac{{\cal H}^{(0)}  +  \delta^{\rm in-in} {\cal H}}{\rho_{\rm bck}^{'(0)} +  \delta^{\rm in-in} \rho_{\rm bck}'} 
 \left[ \rho^0 \left( \vec{k} \right) + \delta^{\rm in-in} \rho \left( \vec{k} \right) \right] \simeq   - \frac{{\cal H}^{(0)}}{\rho_{\rm bck}^{'(0)}} \, V' \,  \left[ \delta \phi^0 \left( \vec{k} \right) + \delta^{\rm in-in} \phi \left( \vec{k} \right) \right] \nonumber\\ 
 &\simeq&   - \frac{{\cal H}^{(0)} \, V'}{- 3 {\cal H}^{(0)} \dot{\varphi}^2} \; \delta \phi \left( \vec{k} \right) = - \frac{H}{\dot{\varphi}} \,  \delta \phi \left( \vec{k} \right) \;, 
  \end{eqnarray} 
where the time dependence of all quantities is understood, and where $H$ denotes the unperturbed physical Hubble rate. 

This concludes the proof that the approximated expression  (\ref{zeta-main}) for $\zeta$ used in the main text is accurate also for computing the effects of particle production. The computation performed in this Appendix neglects metric perturbations. In the next Appendix we show that this is also an accurate approximation, for the purpose of computing the effects of the particle production to leading order in slow roll.

\section{Justification of $\delta g_{\mu \nu} = 0$ } 
\label{app:dg}

In this Appendix, we discuss the effect of metric perturbations and show that, to lowest order, they do not affect the results derived above.  Metric perturbations will generically have three effects: first, they can modify $\varphi$ and the mode functions $\delta \phi_{\vec{k}}$ and $\chi_{\vec{k}}$, secondly, they can modify the relation between $\zeta$ and $\delta \phi$, and finally, they introduce new vertices which can contribute to our calculations of $\left< \delta \phi \, \delta \phi \right>$.

As mentioned above, we will work exclusively in the spatially flat gauge, defined via the line element in \eqref{line}.  In order to avoid confusion between the metric and the coupling constant between the inflaton and $\chi$ fields, we will rename the coupling constant $g$ to $h$ in the first part of  this appendix.

The action can be written as
\begin{align}
S &= S_\phi \left[ \varphi + \delta \phi ,\, g^{(0)} + \delta g \right] + S_{\phi,\chi} \left[  \varphi + \delta \phi ,\, \chi ,\, g^{(0)} + \delta g \right] \;, \nonumber\\ 
S_\phi &=  \int d^4 x \; \sqrt{-g} \; \left[ \frac{M_p^2}{2} \, R - \frac{1}{2} g^{\mu \nu} \partial_\mu \phi \partial_\nu \phi - V \left( \phi \right) \right] \;, 
\nonumber\\ 
 S_{\phi,\chi} &=   \int d^4 x \; \sqrt{-g} \; \left[  - \frac{1}{2} g^{\mu \nu} \partial_\mu \chi \partial_\nu \chi - \frac{h^2}{2} \;  \left( \phi - \phi_* \right)^2 \chi^2 \right] \;, 
\end{align} 
where $S_\phi$ is the part of the action which only involves the inflaton, while $S_{\phi,\chi}$ contains the effect of the $\chi$ fields in addition to the coupling between the inflaton and the $\chi$ fields.

We define our perturbative expansion using the zeroth order action
\begin{align} 
S^{(0)} &\equiv  S_\phi \left[ \varphi + \delta \phi ,\, g^{(0)} + \delta g \right] + S_{\phi,\chi} \left[  \varphi  ,\, \chi ,\, g^{(0)}  \right] \nonumber\\ 
&=   \int d^4 x \; \sqrt{-g} \; \left[ \frac{M_p^2}{2} \, R - \frac{1}{2} g^{\mu \nu} \partial_\mu \phi \partial_\nu \phi - V \left( \phi \right) \right] \nonumber\\ 
& \quad\quad + \int d^4 x  \frac{a^2}{2 } \left[  \chi'{}^2 - \partial_i \chi \, \partial_i \chi - a^2 \, \frac{h^2 }{2} \left( \varphi - \phi_* \right)^2 \chi^2 \right] \,. 
\label{S0-gd}
\end{align} 
In $S_\phi$, we have included the effect of the metric perturbations; however, in our zeroth-order action, we do not include the effect of metric perturbations in $S_{\chi,\phi}$.  We note that by construction, the metric perturbation does not affect $\chi$ in the unperturbed action.  Therefore, at lowest order, the calculation of its unpertured mode functions $\chi^{(0)}$ proceeds as in section \ref{subsec:zero}. On the other hand, the inclusion of metric perturbations in $S^{(0)}$ impacts the unperturbed inflaton mode function, as we discuss below. 

With this definition of the unperturbed action, the interaction terms are given by
\begin{align}
S^{(\rm int)} &=  S_{\phi,\chi} \left[  \varphi + \delta \phi ,\, \chi ,\, g^{(0)} + \delta g \right]  -  S_{\phi,\chi} \left[  \varphi  ,\, \chi ,\, g^{(0)}  \right] \nonumber\\ 
&=  a^4 \Bigg[ -  h^2 \left( \varphi - \phi_* \right) \delta \phi \, \chi^2 - \frac{h^2}{2} \delta \phi^2 \, \chi^2  - \frac{1}{2 a^2} \Phi \chi' \chi' -   \frac{1}{2 a^2} \Phi\, \partial_i \chi\, \partial_i \chi   - \frac{h^2}{2} \, \Phi \, \left( \phi - \phi_* \right)^2 \chi^2  \nonumber \\
& \qquad   - \frac{1}{a^2} \chi' \, \partial_i \chi \, \partial_i \, B  \Bigg] , 
\label{Sint-dg}  
\end{align} 
up to second order in the metric perturbations.  The first two terms give the vertices in $\mathcal{L}_\mathrm{int}$ in \eqref{action}.  The remaining four terms are induced by the metric perturbations and will produce new vertices that we analyze below.  The first three of these terms, involving $\Phi$ (namely, the relativistic generalization of the Newtonian potential), encode the coupling of this metric perturbation to the energy density of the $\chi$ field.

First, we solve for the functions $\Phi$ and $B$ which define the metric perturbations using the zeroth order action.  We still formally decompose the unperturbed mode functions $\delta \phi^{(0)}$ and $\chi^{(0)}$ according to eqs.  \eqref{phi-deco} and \eqref{chi-deco}, and we perform identical  decompositions for $\Phi$ and $B$.  The two metric perturbations $\Phi$ and $B$ are non-dynamical, meaning that they enter in the action without time derivatives. They obey the constraint equations 
\begin{align}
\Phi^{(0)} &= \frac{\varphi'}{2 M_p {\cal H}} \; \frac{\delta \phi^{(0)}}{M_p} \;,  \nonumber\\ 
B^{(0)}  &= {\cal H} \; \Delta^{-1} \left\{ - \frac{\varphi'}{2 {\cal H} M_p} \; \frac{\delta \phi^{'(0)} }{{\cal H} \, M_p} - \frac{1}{2} \left[ \frac{3 \varphi'}{{\cal H} \, M_p} + \frac{a^2 V'}{{\cal H}^2 \, M_p} - \frac{\varphi^{' 3}}{2 {\cal H}^3 M_p^3} \right] \, \frac{\delta \phi^{(0)} }{M_p} \right\} ,
\label{eq:Phi_and_B}
\end{align} 
which are obtained by extremizing the unperturbed action, up to second order in the metric perturbations, with respect to them. (These constraint equations  are known as the hamiltonian and the momentum constraint in the ADM formalism \cite{Arnowitt:1962hi}.) We stress that  the superscript $(0)$ indicates that these are obtained from the unperturbed action.~\footnote{As $\chi^{(0)}$ is not coupled to $\delta g^{(0)}$ nor $\delta \phi^{(0)}$, the relation (\ref{eq:Phi_and_B}) and the computations below up to eq. (\ref{S2-metric}) are a review of the standard computation of linearized primordial perturbations with a single inflaton field, see for instance \cite{Riotto:2002yw} for details.}

We substitute these back into the action.    Using the (unperturbed) slow-roll solution  $\dot{\varphi} =  {\rm sign} \left( \dot{\varphi} \right) \, \sqrt{2 \epsilon} H M_p$, we may write
\begin{align}
\Phi^{(0)} &= \sqrt{\frac{\epsilon}{2}} \; {\rm sign} \left( \dot{\varphi} \right) \; \frac{\delta \phi^{(0)}}{M_p} \;,  \nonumber\\ 
B^{(0)}  &= - \sqrt{\frac{\epsilon}{2}} \; {\rm sign} \left( \dot{\varphi} \right) \;  a \;  H \; \Delta^{-1} \left\{ \frac{\delta \phi^{'(0)} }{a \,  H \, M_p} + \left( \eta - 2 \epsilon \right)    \, \frac{\delta \phi^{(0)} }{M_p} \right\}.
\label{eq:Phi_and_B_2}
\end{align}

Inserting the solutions (\ref{eq:Phi_and_B}) for the nondynamical modes back into the quadratic action for the perturbations we obtain 
\begin{equation}
S_\phi \Big\vert_{\rm second \; order} = \int d t \; d^3 k \; \frac{a^2}{2} \left[ \vert \delta \phi^{'(0)} \vert^2 - \left[ k^2 + a^2 V'' + \frac{3 \varphi^{' 2}}{M_p^2} + \frac{2 a^2 V' \varphi'}{{\cal H} M_p^2} - \frac{\varphi^{' 4}}{2 {\cal H}^2 M_p^4} \right] \vert \delta \phi^{(0)} \vert^2 \right] ,
\end{equation}
which can be rewritten as 
\begin{equation}
S_\phi \Big\vert_{\rm second \; order} = \int d t \; d^3 k \; \frac{a^2}{2} \left[ \vert \delta \phi^{'(0)} \vert^2 - \left[ k^2 + a^2 H^2 \left(3 \eta - 6 \epsilon + {\rm O } \left(  \epsilon^2 \right) \right) \right] \vert \delta \phi^{(0)} \vert^2 \right] .
\label{S2-metric}
\end{equation}

The extremization of this action leads to \cite{Riotto:2002yw}
\begin{equation}
\left( a \delta \phi^{(0)} \right)'' + \left[ k^2 - \frac{1}{\tau^2} \left( 2 + 9 \epsilon - 3 \eta + {\rm O } \left( \epsilon^2 \right) \right) \right] \left( a  \delta \phi^{(0)} \right) = 0 \;, 
\end{equation} 
which is solved by eq. (\ref{df0}) to leading order in slow roll. As metric perturbations do not enter in $  S_{\phi,\chi} $, the unperturbed mode functions $\chi^{(0)}$ are still given by eq. (\ref{dc0}). Namely, the unperturbed solutions coincide with those obtained in  Section \ref{subsec:zero}, up to ${\rm O} \left( \epsilon ,\, \eta \right)$ terms that we can neglect. 

Next, we consider how metric perturbations affect the relation between $\zeta \equiv - {\cal H} \, \frac{\delta \rho}{\rho_{\rm bck}'}$ and $\delta \phi$.  To first order in the metric perturbations, we have 
\begin{align}
\rho = - T^0_0  =  \frac{1}{a^2}& \left[  \frac{1}{2 } \phi' \, \phi' +  \frac{1}{2 } \partial_i \phi\, \partial_i  \phi +  \frac{1}{2 } \chi'\,  \chi' +  \frac{1}{2 } \partial_i \chi \,\partial_i  \chi\right. \nonumber\\
&+ \left.a^2 \, V \left( \phi \right) + a^2 \, \frac{h^2}{2} \left( \phi - \phi_* \right)^2 \chi^2 - \Phi \;  \phi'  \phi'  - \Phi \;  \chi'  \chi'  \right].
\end{align}
The inclusion of metric perturbations introduces the last two terms, proportional to $\Phi$. Perturbations modify the expectation value of the energy density  only to second order, which is an effect that we can safely disregard in our computations. Therefore we can disregard the impact of the metric perturbations in $\rho_{\rm bck}$. Moreover, we see that the metric perturbations correct (in a subdominant way, since we are in a perturbative regime) the kinetic energy of the inflaton and the field $\chi$. As we  saw  in  Appendix \ref{app:zeta}, the two terms  $\phi'{}^{2}$ and $\chi'{}^{2}$  provide  negligible contributions to $\delta \rho$. Therefore, the two corrections $\Phi\, \phi'{}^{2}$ and $\Phi\, \chi'{}^{2}$ are even more negligible. In summary, we can disregard the effect of the metric perturbations on the formal expression of $\zeta$ and continue to take $\zeta \simeq - \frac{\cal H}{\phi'} \, \delta \phi$. 

Therefore, the only possibly relevant effect of the metric perturbations is due to the additional vertices present in (\ref{Sint-dg}). Substituting \eqref{eq:Phi_and_B_2} into the interaction action produces the interaction Hamiltonian, 
\begin{align}
 H_\mathrm{int}(\tau)
&= \int d^3x \, a^4 \Bigg\{ g^2 \left( \phi_0 - \phi_* \right) \delta \phi \, \chi^2 + \frac{g^2}{2} \delta \phi^2 \, \chi^2 \nonumber\\ 
&  \qquad  \qquad  \qquad 
 + \frac{\mathrm{sgn}(\dot{\phi}_0)}{2 a^2} \sqrt{ \dfrac{\epsilon}{2} } \dfrac{\delta \phi}{M_p} \left(\chi^\prime \chi^\prime + \partial_i \chi \partial_i \chi + g^2 a^2 (\phi_0 - \phi_*)^2 \chi^2 \right)  \nonumber \\
&  \qquad  \qquad  \qquad  - \frac{ {\rm sgn} ( \dot{\phi}_0 ) H}{a} \sqrt{\frac{\epsilon}{2}}  \chi' \, \partial_i \chi  \; \Delta^{-1} \left[ \frac{\partial_i \delta \phi^{'} }{a \,  H \, M_p} + \left( \eta - 2 \epsilon \right)    \, \frac{\partial_i \delta \phi }{M_p} \right]  \Bigg\} \,.
\label{eq:Hint_v1}
\end{align}
We note that we have returned to using  $g$ in place of $h$ for the coupling constant between the inflaton and $\phi$, as there is no longer any risk of confusing it with the metric. This interaction Hamiltonian is used in the in-in relation (\ref{P-int}), with the mode functions $\delta \phi$ and $ \delta \chi$  evaluated at the unperturbed level. 

The last two lines in (\ref{eq:Hint_v1}) are  new trilinear $\delta \phi \, \chi^2$ terms generated by the metric perturbations. 
We verified by explicit computations that one loop diagrams with vertices from the last two lines of (\ref{eq:Hint_v1}), namely those emerging from the metric perturbations, are negligible with respect to the dominant one loop diagram obtained from the first line only, which we have computed in the main text. 

As an example, we report the result for the one-loop diagram diagram originating solely from the second line of (\ref{eq:Hint_v1}), namely from interactions due to the $\delta g_{00}$ component.  (We use this result in section \ref{3fld} when we discuss the three-field version of the model.)  We verified that this dominates over diagrams with interactions originated from the third line (namely, from $\delta g_{0i}$), so this is the dominant process that originates from metric perturbations. Diagrammatically, the process is still described by the first loop in Figure \ref{fig:diagrams}, where now the two trilinear vertices are taken from the second line of (\ref{eq:Hint_v1}). The computation follows precisely the same steps that led from eq. (\ref{app-start-explicit-1loop}) to eq. (\ref{app-end-explicit-1loop}), leading to 
\begin{eqnarray}
\delta_{\rm grav} \;  P_\zeta \left( \tau ,\, k \right) \Bigg\vert_{-k \tau \ll 1}  \simeq    \frac{ \epsilon^2 \, H^2 }{k^6} \; \frac{2+\sqrt{2}}{16 \pi^3} \; a_*^3 \; g^{7/2} \; \vert \dot{\varphi}_* \vert^{3/2} 
\left[ \; \int_0^{x_*} d x_1 \, \frac{  x_1 \, \cos \left( x_1 \right) -  \sin \left( x_1 \right)  }{ x_1 }  \; {\rm ln } \left( \frac{x_*}{x_1} \right) \right]^2 , \nonumber\\ 
\end{eqnarray} 
where the suffix ``grav'' indicates that this is the dominant term originating from metric perturbations (namely this is a purely gravitational interaction between the $\chi-$quanta and the inflaton field), and where we recall that $x_* = \frac{k}{a_* \, H}$. The logarithm in the integral originates from a factor $\omega \simeq a\, \vert \varphi \left( \tau_1 \right) - \phi_* \vert$ present in this interaction, and from the approximation (\ref{mass-app}).

Performing the integration and dividing by the unperturbed power spectrum, we obtain 
\begin{eqnarray} 
\frac{\delta_{\rm grav} P_\phi}{P_\phi^{(0)}} & \simeq &   \frac{2+\sqrt{2}}{8 \pi^3} \;  \frac{  \epsilon^2  \, g^{7/2} \; \vert \dot{\varphi}_* \vert^{3/2} }{H^3} \, 
{\cal S}_{\rm grav} \left[ x_* \right] \;, \nonumber\\ 
{\cal S}_{\rm grav} \left[ x_* \right] &\equiv& \frac{1}{x_*^3} \left[   - x_* \; {\rm Re } \left\{ _3F_3 \left(1,\, 1 ,\, 1;\,  2,\, 2 ,\, 2 ;\, i \, x_* \right)\right\} + {\rm SinIntegral } \left( x_* \right)  \right]^2  \;. 
\label{dP2-grav-full}
\end{eqnarray} 
The shape function behaves as $\frac{x_*^3}{729}$ at small $x_*$, and as $ \frac{\pi^2 \left( \ln x_* + \gamma - 1 \right)^2}{4 x_*^3} $ at large $x_*$. As the result studied in the main text, it is characterized by a peak at the scales that exited the horizon during the production of the $\chi-$quanta. The maximum value of this peak gives 
\begin{eqnarray} 
\frac{\delta P_\phi}{P_\phi^{(0)}} \Big\vert_{\rm peak} & \simeq & 4 \cdot 10^{-4} \,  \frac{  \epsilon^2  \, g^{7/2} \; \vert \dot{\varphi}_* \vert^{3/2} }{H^3}  \;\;\; {\rm at } \;\;\; \frac{k}{a_* \, H} \simeq 4.67 \;, 
\label{dP2-grav}
\end{eqnarray} 
which, as anticipated, is indeed subdominant with respect to the result (\ref{dPZ-peak}), which originated from the direct coupling in the potential. 

We now discuss the reason why the interactions due to the metric perturbations provide subdominant contributions with respect to the direct coupling of the potential,  by estimating the magnitudes of the various  vertices entering in (\ref{eq:Hint_v1}). In the adiabatic regime (namely, once the production of quanta of $\chi$ is saturated, and the expression  (\ref{dc0}) is valid), we have $\chi^{(0)'} \sim \omega\, \chi^{(0)} \simeq g \, a \, \vert \phi_0 - \phi_* \vert \,  \chi^{(0)}  \gg \partial_i \chi^{(0)}$. Using this approximation in the last two lines of eq. (\ref{eq:Hint_v1}) gives (we note that, with this approximation, in the third line we have $\chi^{(0)'} \partial_i \chi^{(0)} \simeq \frac{\omega}{2} \partial_i \chi^{(0)2}$, and we then moved the spatial derivative on $\delta \phi^{(0)}$) 
\begin{align}
 H_\mathrm{int}(\tau)
&\sim \int d^3x \, a^4 \Bigg\{ g \, \frac{\omega}{a} \delta \phi^{(0)} \, \chi^{(0)2} + \frac{g^2}{2} \delta \phi^{(0)2} \, \chi^{(0)2} \nonumber\\ 
&  \qquad  \qquad  \qquad 
 + \mathrm{sgn}(\dot{\phi}_0)  \, \sqrt{\frac{\epsilon}{2}} \,   \frac{\omega}{a \, M_p} \, \frac{\omega}{a} \, \delta \phi^{(0)} \, \chi^{(0)2} \nonumber\\ 
 &  \qquad  \qquad  \qquad  +  {\rm sgn} ( \dot{\phi}_0 )  \sqrt{\frac{\epsilon}{2}}   \frac{H}{2 M_p}   \;  \left[ 
  - \frac{k^2 \tau^2}{1+i k \, \tau}  + \left( \eta - 2 \epsilon \right)   \right] \frac{\omega}{a} \delta \phi^{(0)} \,  \chi^{(0)2}  \Bigg\} \,, 
\end{align}
where expression (\ref{df0}) has been used in the third line. In comparing the three $\delta \phi  \chi^2$ vertices, we need to compare 
\begin{equation}
g \;\;\;\; {\rm vs.} \;\;\;\;  \sqrt{\frac{\epsilon}{2}} \,   \frac{\omega}{a \, M_p} \;\;\;\; {\rm vs.} \;\;\;\;  \sqrt{\frac{\epsilon}{2}}   \frac{H}{2 M_p}   \;  \left[ 
  - \frac{k^2 \tau^2}{1+i k \, \tau}  + \left( \eta - 2 \epsilon \right)   \right] \;, 
\end{equation}
which are, respectively, the contributions from the direct coupling in the potential (the one considered in the main text), from the metric perturbation $\Phi$, and from the metric perturbation $B$. It is immediate to see that the first term dominates. 

We also note that the ratio between the cubic gravitational vertex involving $\Phi$ and the vertex the arises directly from the potential parametrically amounts to $\sim \sqrt{\epsilon} \, \frac{\vert \varphi - \phi_*\vert}{M_p} \sim  \sqrt{\epsilon} \frac{ \vert \dot{\varphi}_* \vert }{H \, M_p} \sim \epsilon$ (where in the second step the approximation (\ref{mass-app}) has been used). This explains why the result  (\ref{dP2-grav}) is ${\rm O} \left( \epsilon^2 \right)$ suppressed with respect to (\ref{dPZ-peak}). 

This concludes the proof that the metric perturbations provide a negligible correction to the correction of the scalar power spectrum due to particle production.

\section{Dominant contribution in the three field model } 
\label{app:3fld-dominant}

In this Appendix, we evaluate the contribution to the scalar power spectrum given by the second diagram in Figure \ref{fig:3fld-diag}. This process corresponds to four insertions of the interaction hamiltonian in the in-in formalism, 
\begin{eqnarray} 
&& \delta \left\langle \delta \phi \left( \tau ,\, \vec{k} \right)  \delta \phi \left( \tau ,\, \vec{k}' \right)   \right\rangle = \left( - i \right)^4 \int^\tau d \tau_1  \int^{\tau_1} d \tau_2  \int^{\tau_2} d \tau_3    \int^{\tau_3} d \tau_4   \nonumber\\ 
&& \quad\quad\quad\quad  \times 
\left\langle \left[  \left[  \left[ \left[ \delta \phi^{(0)} \left( \tau ,\, \vec{k} \right)  \delta \phi^{(0)} \left( \tau ,\, \vec{k}' \right)  ,\, H_{\rm int}^{(0)} \left( \tau_1 \right) \right] ,\,  H_{\rm int}^{(0)} \left( \tau_2 \right) \right] ,\,  H_{\rm int}^{(0)} \left( \tau_3 \right) \right] 
 ,\,  H_{\rm int}^{(0)} \left( \tau_4 \right) \right]  \right\rangle \;,  \nonumber\\ 
\end{eqnarray} 
where two of the interaction hamiltonians are given by the first term in (\ref{Hint-3fld}), and the other two by the third term in  (\ref{Hint-3fld}). 
We work out the commutations and the expectation value in this diagram as we did for the other processes, see for instance Appendix \ref{app:PS}. After some lengthy algebra, we obtain 
\begin{eqnarray} 
&& 
\delta \left\langle \delta \phi \left( \tau ,\, \vec{k} \right)  \delta \phi \left( \tau ,\, \vec{k}' \right)   \right\rangle' 
\simeq  \frac{9 \left( 2 + \sqrt{2} \right)}{4 \pi^3}    \,   h^{7/2} \, \vert \dot{\psi}_* \vert^{3/2} \,  \epsilon_\phi \, \epsilon_\psi \, 
\frac{1}{H \, k^3  x_*^3} \; \nonumber\\ 
&& \quad\quad  \quad\quad  \quad\quad  \quad\quad  \quad\quad  \quad\quad  \quad\quad  \quad\quad 
 \times 
\left\{ \int_x^{x_*} \frac{d x_1}{x_1^4} \, \left[ x_1 \cos \left( x_1 \right) - \sin \left( x_1 \right) \right] \; {\cal F} \left[ x_1 ,\, x_* \right] \right\}^2 \;, \nonumber\\ 
\label{3fld-dominant-par}
\end{eqnarray} 
where 
\begin{eqnarray} 
 && 
 {\cal F} \left[ x_1 ,\, x_* \right] \equiv \int_{x_1}^{x_*} \frac{d x_2}{x_2} \,   
\left[ - \left( x_1 -  x_2 \right) \cos  \left( x_1 - x_2 \right) + \left( 1 + x_1 \, x_2 \right)  \sin  \left( x_1 - x_2 \right) \right] \nonumber\\ 
 &&   = 
x_1 \left[ -1 + \cos \left( x_1 - x_* \right) \right] + {\rm CI} \left( x_1 \right) \left[ x_1 \cos \left( x_1 \right) - \sin \left( x_1 \right) \right]  +  {\rm CI} \left( x_* \right) \left[ - x_1 \cos \left( x_1 \right) + \sin \left( x_1 \right) \right]  \nonumber\\ 
&& - \sin \left( x_1 - x_* \right) 
+ \left[ \cos \left( x_1 \right) + x_1 \sin \left( x_1 \right) \right] \left[ {\rm SI} \left( x_1 \right) -  {\rm SI} \left( x_* \right) \right] \;,
\end{eqnarray} 
(with CI and SI denoting, respectively, the cosine and sine integral). In these relations $x \equiv - k \, \tau$ and $x_* \equiv - k \tau_*$ where $\tau_*$ is the moment at which the particle production takes place. The (rescaled) time integral giving ${\cal F} \left[ x_1 ,\, x_* \right]$ refers to the innermost part of the diagram, and it gives the $\delta \psi$ mode at the time $\tau_1= -x_1 / k$ due to the particle production that occurred at $\tau_*$. The (rescaled) time integral in (\ref{3fld-dominant-par}) is due to the two other legs in the diagram, and  it encodes the $\delta \psi \rightarrow \delta \phi$ transition  that continuously takes place between $\tau_*$ and $\tau$, due to the quadratic $\delta \phi \, \delta \psi$ term in  (\ref{Hint-3fld}). 

We divide (\ref{3fld-dominant-par}) by the zeroth order correlator $ \left\langle \delta \phi^{(0)} \left( \tau ,\, \vec{k} \right)  \delta \phi^{(0)} \left( \tau ,\, \vec{k}' \right)   \right\rangle'  \simeq \frac{H^2}{2 k^3} $. This ratio gives the correction $\delta P / P^{(0)}$ to the power spectrum. The remaining integral in (\ref{3fld-dominant-par}) is dominated by the smallest values of $x_1$, where we can expand it and obtain 
\begin{eqnarray}
\frac{\delta P_\phi}{P_\phi^{(0)}} &\simeq& \frac{9 \left( 2 + \sqrt{2} \right)}{2 \pi^3}    \, \frac{  h^{7/2} \, \vert \dot{\psi}_* \vert^{3/2} }{ H^3 }  \,  \epsilon_\phi \, \epsilon_\psi \, \frac{1}{   x_*^3} \; \left\{ \frac{- \sin \left( x_* \right) + {\rm SI} \left( x_* \right)}{3} \,  \int_x^{x_*} \frac{d x_1}{x_1} \right\}^2 \nonumber\\ 
&\simeq& \frac{ \left( 2 + \sqrt{2} \right)}{2 \pi^3}    \, \frac{  h^{7/2} \, \vert \dot{\psi}_* \vert^{3/2} }{ H^3 }  \,  \epsilon_\phi \, \epsilon_\psi \, N_k^2 \; 
\frac{\left[ - \sin \left( x_* \right) + {\rm SI} \left( x_* \right) \right]^2}{   x_*^3} \;. 
\label{3fld-dominant-res}
\end{eqnarray}
where $N_k$ is the number of e-folds during which the mode $x$ is outside the horizon. To be more precise, the result (\ref{3fld-dominant-res}) 
assumes $\epsilon_\psi \propto \dot{\psi}_*^2 = {\rm constant}$ all throughout inflation. If instead $\psi$ gets stabilized at some time before the end of inflation, 
$N_k$ denotes the number of folds between the moment the modes exits the horizon and the moment that $\psi$ gets stabilized.

\end{document}